\documentclass[12pt]{JHEP3}

\usepackage[dvips]{graphicx}
\usepackage{axodraw}
\usepackage{subfig}
\usepackage{amsmath, amssymb}
\usepackage{url}
\usepackage{float}

\newcommand{\ie}{\textit{i.e.}}
\newcommand{\etc}{\textit{etc}}
\newcommand{\figref}[1]{figure~\ref{#1}}
\newcommand{\appref}[1]{appendix~\ref{#1}}
\newcommand{\secref}[1]{section~\ref{#1}}
\newcommand{\mcap}[1]{\text{\footnotesize #1}}
\newcommand{\MHVb}{$\overline{\text{MHV}}$}

\DeclareMathOperator*{\Res}{Res}
\def\Rec{R}

 \usepackage{cite}

\newskip\humongous \humongous=0pt plus 1000pt minus 100pt
\def\caja{\mathsurround=0pt} \def\eqalign#1{\,\vcenter{\openup1\jot
    \caja \ialign{\strut \hfil$\displaystyle{##}$&$
      \displaystyle{{}##}$\hfil\crcr#1\crcr}}\,} \newif\ifdtup

\def\oneloop{\text{1-loop}} 
\def\tree{\text{tree}}

\def\la {\langle} \def\ra{\rangle}

\def\spa#1.#2{\left\langle#1\,#2\right\rangle}
\def\spb#1.#2{\left[#1\,#2\right]}
\def\spba#1.#2.#3{\left[#1|#2|#3\right\rangle}
\def\spab#1.#2.#3{\left\langle#1|#2|#3\right]}
\def\spaa#1.#2.#3{\left\langle#1|#2|#3\right\rangle}\def\spbb#1.#2.#3{\left[#1|#2|#3\right]}
\def\lor#1.#2{\left(#1\,#2\right)} 

\def\YM{Yang--Mills\ } 
\def\NeqEight{{\cal N}=8}

\def\mpppp{1^-,2^+,3^+,4^+,5^+}

\newcommand{\theeqnumber}{\thesection.\arabic{equation}}
\def\equn{\refstepcounter{equation} \eqno({\rm \theeqnumber}) }

\author{David C. Dunbar, James H. Ettle and Warren B. Perkins
  \\ Department of Physics,\\
  Swansea University,\\
  Swansea, SA2 8PP, UK\\
E-mail: \email{d.c.dunbar@swan.ac.uk},
\email{j.h.ettle@swan.ac.uk}, \email{w.perkins@swan.ac.uk} }

\abstract{We present a semi-recursive method for calculating the rational parts of one-loop
  gravity amplitudes which utilises axial gauge diagrams to determine
  the non-factorising pieces of the amplitude.  This method is used to compute the amplitudes 
  $M^\oneloop(1^-,2^+,3^+,4^+,5^+)$ and
  $M^\oneloop(1^-,2^+,3^+,4^+,5^+,6^+)$.}


\title{Augmented Recursion For One-loop Gravity Amplitudes}
\begin{document}

\section{Introduction}

On-shell recursive techniques have proven very successful in the
computation of scattering amplitudes in gauge theories and in theories
of gravity \cite{Britto:new,BBSTgravity,CSgravity}.  The recursive
techniques for tree scattering amplitudes make use of both the
rationality of the amplitudes and their complex factorisation
properties.
Specifically, in a theory with massless states, if we use a spinor
helicity representation for the polarisation vectors it is possible to
write the amplitude entirely in terms of spinorial variables
$A(\lambda^i_\alpha,\bar\lambda^i_{\dot \alpha})$ where the massless
momentum of the $i^\text{th}$ particle is 
$\lambda_\alpha^i\bar\lambda_{\dot \alpha}^i=(\sigma_\mu)_{\alpha\dot
  \alpha} k^\mu_i$.\footnote{We use the usual spinor products 
  $\spa{a}.{b}= \epsilon^{\alpha\beta}\lambda_\alpha^a\lambda_\beta^b$,
  $\spb{a}.{b}= \epsilon^{\dot\alpha\dot\beta}\bar\lambda_{\dot\alpha}^a\bar\lambda_{\dot\beta}^b$,
  which satisfy $\spa{a}.b\spb{b}.{a} =(k_a+k_b)^2\equiv s_{ab}$, chains of spinor products such as
  $[a|b|c\ra\equiv \spb{a}.b\spa{b}.c$ and $[a|P_{ef\cdots}|b\ra=[a|e|b\ra+[a|f|b\ra+\cdots$ etc and $t_{abc}\equiv (k_a+k_b+k_c)^2$. } 
   
Within this formalism it is possible to probe the analytic
structure of the amplitude by choosing a pair $a,b$ of external momenta and
shifting these according to
$$
\bar\lambda^a \longrightarrow \bar\lambda^a-z\bar\lambda^b,\qquad 
    \lambda^b \longrightarrow     \lambda^b+z    \lambda^a \equn
\label{eqshift}
$$
where we suppress the spinor indices. The analytic
behaviour of the shifted amplitude $A(z)$ can then be studied.  

\break 
\noindent
If $A(z)$
\begin{enumerate}
\item is a rational function,
\item \label{recreq2} has simple poles at points $z_i$, and
\item \label{recreq3} vanishes as $z\longrightarrow \infty$
\end{enumerate}
then applying Cauchy's theorem to $A(z)/z$ with a contour at
infinity yields
$$
A(0)=-\sum_{z_i} \Res\biggl( {A(z)\over z}, z_i \biggr).
\equn\label{EqCauchyRes}
$$
This technique has proven very effective in computing tree amplitudes
and has been extended from the purely gluonic case to a variety of
other applications including that of
gravity~\cite{BBSTgravity,CSgravity}.  Alternate shifts
exist~\cite{Risager:2005vk} which can be used to re-derive the CSW
formulation for Yang--Mills~\cite{Cachazo:2004kj} and
gravity~\cite{BjerrumBohr:2005jr}.

The result~(\ref{EqCauchyRes})
holds even if condition~\ref{recreq2} above is relaxed to poles of finite
order; however this condition allows us to use the factorisation
theorems to determine the residues in terms of lower point amplitudes.
At tree level the factorisation is relatively simple:
amplitudes must factorise on multi-particle and collinear poles. For a
partition of the external momenta $(S_L,S_R)$
with at least two momenta on either side,
and defining
$K^\mu \equiv \sum_{i\in S_L} k_i^\mu$, the $n$-point tree amplitude
$A_n^\tree$ factorises as $K$ becomes on shell as
$$
A_{n}^{\tree}\ \mathop{\longrightarrow}^{K^2 \rightarrow 0}
\sum_{\sigma} \Biggl[ A_{r+1}^{\tree}\big(k_i\in S_L , K^\sigma\big)
\, {i \over K^2} \, A_{n-r+1}^{\tree}\big((-K)^{-\sigma}, k_i \in S_R
\big) \Biggr] \equn
$$ where $\sigma$ denotes the internal state of the intermediate 
particle and $r$ is the length of $S_L$.
Consequently, simple poles in the shifted amplitude $A(z)$ occur at
values of $z$ where $K^2(z)=0$.  Only those $K$'s containing precisely one
of $k_a$ or $k_b$ will be $z$ dependent. When the corresponding
$K^2(z)$ vanishes the residue will be the product of the tree
amplitudes defined at $z=z_i$. Thus we can express the $n$-point tree
amplitude in terms of lower point amplitudes,
$$
A_n^\tree (0) \; = \; \sum_{i,\sigma} {A^{\tree,\sigma}_{r_i+1}(z_i)
  {i\over K^2}A^{\tree,-\sigma}_{n-r_i+1}(z_i)},
\equn\label{RecursionTree}
$$
where the summation over $i$ is only over factorisations where the $a$
and $b$ legs are on opposite sides of the pole.

Beyond tree level there are three potential barriers to using recursion. Firstly
the amplitudes, in general, contain non-rational functions such as
logarithms and dilogarithms; secondly, the amplitudes may contain
higher-order poles for complex momenta and, finally, the amplitudes may not vanish asymptotically with $z$.  
Nonetheless a variety of
techniques based upon recursion and unitarity have been developed.  A
one-loop amplitude for massless particles may be expressed as
$$
A^{\oneloop} = \sum_{n=2,3,4;i} c_i I^i_{n} +R +O(\epsilon) \equn
$$
where the scalar integral functions $I^i_{n}$ are the various scalar
box, triangle and bubble functions.  The function $R$ contains the
remaining rational terms.  The one-loop amplitude can then be
specified by computing the coefficients, $c_i$, and the purely rational term $R$.  The
$c_i$ are rational coefficients which can be computed by various applications of
the four-dimensional unitarity 
technique~\cite{BDDKa,Britto:2004nc,Dunbar:2009ax} or indeed recursively~\cite{Bern:2005hh}.

There are a variety of strategies for evaluating the rational
terms. They may be evaluated using $D$-dimensional unitarity, by
recursion or by specialised Feynman diagram 
techniques~\cite{Berger:2006ci, DunitarityA, DunitarityB,
  DunitarityC, DunitarityD, DunitarityE, DunitarityF, blackhat,
  Britto:2004ap, Xiao:2006vt, Binoth:2006hk, Badger:2008cm, OPP, Bern:1996ja}.   
In general, the rational term $R$ does not simply satisfy
the
previously-stated
requisites for recursion \ref{recreq2} and \ref{recreq3}.  If the
amplitude has only simple poles but does not vanish as
$z\longrightarrow \infty$ then it can be possible to formulate
recursion by the use of an auxiliary recursion
relation~\cite{Berger:2006cz}.  However there are rational amplitudes
for which one cannot find a shift which only generates simple
poles such as the one-minus amplitude
$A^\oneloop(1^-,2^+,\cdots,n^+)$.  These amplitudes vanish at tree
level and consequently are purely rational at one-loop.  A shift on
these amplitudes yields double and single poles
$$
A \sim {a \over (z-z_i)^2} +{ b\over (z-z_i) }+\cdots = {a \over
  (z-z_i)^2}\biggl(1 +{b\over a} (z-z_i) +\cdots \biggr) \equn
$$
The double pole is {\it not} in itself a a barrier to using recursion
with the double pole contributing
$$
-\Res\left( { 1 \over z(z-z_i)^2 },z_i \right)={1\over z_i^2}. 
\equn\label{eqDoublePoleResidue}
$$
However to obtain the full residue
in a recursive construction
one must have specific formulae for this double pole and for the
coincident
single pole, or the `pole under the double pole'.

In ref.~\cite{Bern:2005hs} the form of the pole in Yang--Mills was
postulated to be
$$
{1 \over (K^2)^2 } \left( 1+ \sum_{a_i,b_i} S(a_1,\hat K^+,a_2)
  \,K^2\, S(b_1,\hat K^-, b_2) \right) \equn\label{eqYangMillsSoft}
$$
where the `soft' factors are
$$
S(a, s^+, b)= {\spa{a}.{b} \over \spa{a}.s\spa{s}.b } ,\qquad
S(a, s^-, b)= {\spb{a}.{b} \over \spb{a}.s\spb{s}.b } \equn
$$ 
With this ansatz recursion correctly reproduces the known 
one-minus one-loop amplitudes.  In ref.~\cite{Vaman:2008rr} 
it was shown that the consistency requirements for recursion in QCD are sufficient to determine these soft factors.

The above postulate, or variations thereof, however does not work for
gravity amplitudes~\cite{Brandhuber:2007up}.  In this article we will
demonstrate how to apply recursion techniques in gravity scattering
amplitudes by determining the `pole under the pole' using an
axial gauge formalism.  By only keeping the pole terms it is relatively simple to extract these from the diagrammatic approach.
We demonstrate this by calculating the
previously-unknown amplitudes $M^\oneloop(1^-,2^+,3^+,4^+,5^+)$ and
$M(1^-,2^+,3^+,4^+,5^+,6^+)$.
We assume that the shifted amplitudes have vanishing behaviour as $z\longrightarrow\infty$. 
The expressions we derive have the
correct symmetries and soft limits, providing strong evidence for the validity of this assumption.
Further, we compare the
result numerically with a completely independent computation of
$M^\oneloop(1^-,2^+,3^+,4^+,5^+)$ from `string-based rules' for
gravity~\cite{Bern:1991aq,Bern:1991an,Bern:1993wt,Dunbar:1994bn}.

\section{Recursion}
The factorisation of one-loop massless amplitudes is described
in ref.~\cite{BernChalmers},
\begin{equation}\begin{split}
    \label{LoopFact}
    &A_{n}^{\oneloop} \mathop{\longrightarrow}^{K^2 \rightarrow 0}
    \sum_{\lambda=\pm} \Biggl[ A_{r+1}^{\oneloop}\big(k_i, \ldots,
    k_{i+r-1}, K^\lambda\big) \, {i \over K^2} \,
    A_{n-r+1}^{\tree}\big((-K)^{-\lambda}, k_{i+r}, \ldots,
    k_{i-1}\big) \\ & + A_{r+1}^{\tree}\big(k_i, \ldots, k_{i+r-1},
    K^\lambda\big) {i\over K^2}
    A_{n-r+1}^{\oneloop}\big((-K)^{-\lambda}, k_{i+r}, \ldots,
    k_{i-1}\big) \\
    & + A_{r+1}^{\tree}\big(k_i, \ldots, k_{i+r-1}, K^\lambda\big)
    {i\over K^2} A_{n-r+1}^{\tree}\big((-K)^{-\lambda}, k_{i+r},
    \ldots, k_{i-1}\big) F_n\big(K^2;k_1, \ldots, k_n\big) \Biggr],
  \end{split}\end{equation}
where the one-loop `factorisation function' $F_n$ is
helicity-independent.
Na\"ively this only contains single poles, however for
complex momenta there are double poles. These can be interpreted as due to the three-point all-plus (or all-minus) one-loop
amplitude also containing a pole
$$
A^{\oneloop}_3( K^+, a^+,b^+) = { 1 \over K^2 } V^\oneloop(K^+,a^+,b^+)
\equn
$$
where, for pure Yang--Mills,
$$
V^\oneloop(K^+,a^+,b^+) =-{ i \over 48 \pi^2
}\spb{K}.a\spb{a}.b\spb{b}.K   .  
\equn
$$

To see this explicitly, let us consider the five-point Yang--Mills
amplitude \break $A^\oneloop(1^-,2^+,3^+,4^+,5^+)$~\cite{Bern:1993mq}:
$$  
\eqalign{
  A_{5}^\oneloop (1^-, 2^+, 3^+, 4^+, 5^+) =&    
\cr 
{i \over 48 \pi^2} \, {1\over \spa3.4^2} \biggl[
-{\spb2.5^3 \over
      \spb1.2 \spb5.1} +  & {\spa1.4^3 \spb4.5 \spa3.5 \over \spa1.2
      \spa2.3 \spa4.5^2} - {\spa1.3^3 \spb3.2 \spa4.2 \over \spa1.5
      \spa5.4 \spa3.2^2}
      \biggr] .
\cr}      
\equn
$$
If we carry out a complex shift on legs $1$ and $5$,
$$
\lambda^5 \longrightarrow \lambda^5+z\lambda^1 , \qquad 
\bar\lambda^1
\longrightarrow \bar\lambda^1-z\bar\lambda^5 ,
\equn\label{eq:shift-mpppp}
$$
then $\spa4.5 \longrightarrow \spa4.5+z\spa4.1$ which vanishes at $z=
-{\spa4.5/ \spa4.1}$ and the amplitude has a double pole at this
point.\footnote{The term which gives rise to the double pole
  $\spb{4}.5/\spa4.5^2$ is the one-loop splitting
  function~\cite{BDDKa} which only gives a linear collinear pole
  for real momenta.}

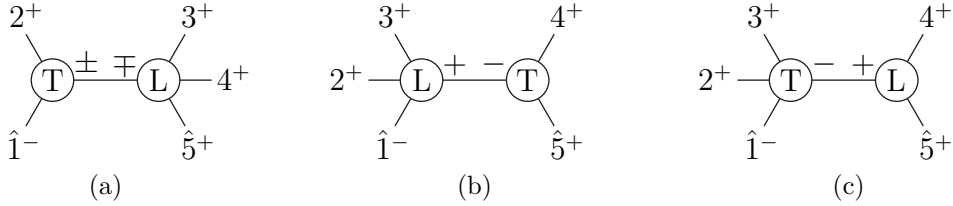
\begin{figure}[h!]
  \begin{center}
    \subfloat[]{
      \begin{picture}(120,60)
        \SetOffset(60,30)
        \Line(-20,0)(20,0)
        \Line(-20,0)(-30,17.3205)
        \Line(-20,0)(-30,-17.3205)
        \Line(20,0)(30,17.3205)
        \Line(20,0)(30,-17.3205)
        \Line(20,0)(40,0)
        \BCirc(20,0){8} \Text(20,0)[cc]{L}
        \BCirc(-20,0){8} \Text(-20,0)[cc]{T}
        \Text(-31,-19)[tc]{$\hat1^-$}
        \Text(-30,20)[bc]{$2^+$}
        \Text(35,20)[bc]{$3^+$}
        \Text(42,1)[lc]{$4^+$}
        \Text(35,-19)[tc]{$\hat5^+$}
        \Text(-12,2)[bl]{$\pm$}
        \Text(12,2)[br]{$\mp$}
      \end{picture}
    }\quad\subfloat[]{
      \begin{picture}(120,60)
        \SetOffset(60,30)
        \Line(-20,0)(20,0)
        \Line(-20,0)(-30,17.3205)
        \Line(-20,0)(-30,-17.3205)
        \Line(20,0)(30,17.3205)
        \Line(20,0)(30,-17.3205)
        \Line(-20,0)(-40,0)
        \BCirc(-20,0){8} \Text(-20,0)[cc]{L}
        \BCirc(20,0){8} \Text(20,0)[cc]{T}
        \Text(-31,-19)[tc]{$\hat1^-$}
        \Text(-42,1)[rc]{$2^+$}
        \Text(-30,20)[bc]{$3^+$}
        \Text(35,20)[bc]{$4^+$}
        \Text(35,-19)[tc]{$\hat5^+$}
        \Text(-12,2)[bl]{$+$}
        \Text(12,2)[br]{$-$}
      \end{picture}
    }\quad\subfloat[]{
      \begin{picture}(120,60)
        \SetOffset(60,30)
        \Line(-20,0)(20,0)
        \Line(-20,0)(-30,17.3205)
        \Line(-20,0)(-30,-17.3205)
        \Line(20,0)(30,17.3205)
        \Line(20,0)(30,-17.3205)
        \Line(-20,0)(-40,0)
        \BCirc(20,0){8} \Text(20,0)[cc]{L}
        \BCirc(-20,0){8} \Text(-20,0)[cc]{T}
        \Text(-31,-19)[tc]{$\hat1^-$}
        \Text(-42,1)[rc]{$2^+$}
        \Text(-30,20)[bc]{$3^+$}
        \Text(35,20)[bc]{$4^+$}
        \Text(35,-19)[tc]{$\hat5^+$}
        \Text(-12,2)[bl]{$-$}
        \Text(12,2)[br]{$+$}
      \end{picture}
      \label{fig:Ampppp-recursion-dblpole}
    }
  \end{center}
  \caption{Diagrams contributing to the recursive construction of
    $A(1^-, 2^+, 3^+, 4^+, 5^+)$ with legs $1$ and $5$ shifted in the
    manner of (2.5). The diagram
    (c) contains the one-loop vertex
    $V^\oneloop(\hat K^+, 4^+, \hat 5^+)$ that contributes the
    double-pole.}
  \label{fig:Ampppp-recursion}
\end{figure}

A recursive approach suggests drawing the diagrams shown in
\figref{fig:Ampppp-recursion}. The third of these involves the one-loop vertex
$V^{\oneloop}(K^+,4^+,\hat 5^+)$. Computing with this does correctly
generate the double pole in the amplitude
\cite{Bern:2005hs,Brandhuber:2007up}, however it needs augmentation to
give an expression with the correct single pole. By trial
and error, adding the second term in \eqref{eqYangMillsSoft} gives the
correct single pole and completes the computation of the amplitude.

When calculating the gravity amplitude 
$M_{5}^\oneloop (1^-, 2^+, 3^+,4^+, 5^+)$ we must consider the same class of diagrams as in
\figref{fig:Ampppp-recursion} together with permutations over the
external legs.  For gravity the vertex
$$
V^{\oneloop}(K^+,a^+,b^+) =-{ i \kappa^3 \over 1440 \pi^2 }(
\spb{K}.a\spb{a}.b\spb{b}.K )^2 \equn
$$
can be used to generate a double pole term which has the correct soft
and collinear limit, but attempts~\cite{Brandhuber:2007up} to
implement a universal correction for the single pole analogous to that
of \eqref{eqYangMillsSoft} have failed.

\begin{figure}[H]
  \centerline{
    \begin{picture}(92,71)
      \SetOffset(-20,-4)
      \ArrowLine(80,60)(80,20)
      \ArrowLine(65,46)(80,60) \ArrowLine(80,20)(65,34)
      \CArc(60,40)(8,0,360) \Line(95,10)(80,20) \Line(95,70)(80,60)
      \Line(55,46)(40,60) \Line(55,34)(40,20) \Line(52,40)(40,40)
      \Text(60,40)[c]{$T$} \Text(74,19)[r]{${}^{l-{\hat 5}}$}
      \Text(74,57)[r]{${}^{l+4}$} \Text(85,40)[c]{$l$}
      \Text(105,70)[c]{$4^+$} \Text(105,10)[c]{${\hat 5}^+$}
      \Text(28,60)[c]{$3^+$} \Text(28,40)[c]{$2^+$}
      \Text(28,20)[c]{${\hat 1}^-$}
    \end{picture}
  }
  \caption{The form of the tree insertion that augments the recursion in
    order to construct the double pole and its underlying single
    pole.}
  \label{fig:tree-insertion}
\end{figure}
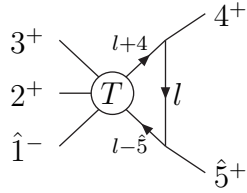

We find that the resolution is to 
replace
the factorisation term of
\figref{fig:Ampppp-recursion-dblpole} with a tree insertion diagram of
the form shown in \figref{fig:tree-insertion} and
compute this using axial gauge diagrammatics.  In
\secref{sec:axialdiags} we present the axial gauge rules, in
\secref{sec:grav-mpppp} the computation of the five point
one-minus gravity amplitude and in \appref{app-sixpoint} the result
of the computation of the six point one-minus amplitude.

\section{Axial gauge diagrammatics}
\label{sec:axialdiags}
\newcommand{\Tau}{\tau}

We use axial gauge diagrammatic methods to determine the singular
structure necessary to augment the recursion.  We identify and compute
the singularities arising when we shift a negative-helicity leg
$a$ and a positive-helicity leg $b$ as in \eqref{eqshift}.  These singularities arise from
propagators involving just external momenta and from the loop momentum
integration.

Following ref.~\cite{Schwinn:2005pi} we use a set of Feynman rules for
Yang--Mills amplitudes based on scalar propagators connecting three
and four point vertices.  The starting point is the expansion of the
axial gauge propagator in terms of polarisation vectors,
$$
\eqalign{ i{d_{\mu\nu}\over k^2} &= {i\over k^2}\left(-g_{\mu\nu}
    +2{k_\mu q_\nu+q_\mu k_\nu \over 2k\cdot q}\right) \cr &={i\over
    k^2}
  [\epsilon_\mu^+(k)\epsilon_\nu^-(k)+\epsilon_\mu^-(k)\epsilon_\nu^+(k)+
  \epsilon_\mu^0(k)\epsilon_\nu^0(k) ], \cr}
\label{propa}
\equn
$$
where \newcommand{\nulled}{\flat}
$$
\epsilon^+_\mu(k) = {[k^\nulled | \gamma_\mu | q \ra \over \sqrt{2}
  \spa{k^\nulled}.q}, \qquad \epsilon^-_\mu(k)= { [
  q|{\gamma_\mu}|{k^\nulled} \ra \over \sqrt{2} \spb{k^\nulled}.q}, \qquad
\epsilon^0_\mu(k)=2{\sqrt{k^2}\over 2k\cdot q}q_\mu.  \equn
$$
Here $q$ is a null reference momentum which may be complex. For any
momentum $k$ we define its \emph{$q$-nullified} form
$$
k^\nulled := k - {k^2\over 2k\cdot q} q.  \equn\label{eqqnullification}
$$
Contracting the polarisation vectors into the usual Yang--Mills
three-point vertex yields the familiar three-point MHV and ${\overline
  {\rm MHV}}$ vertices,
$$
\eqalign{ {1\over i\sqrt{2}}V_3(1^-,2^-,3^+)&={{\spa 1.2}^3\over \spa
    2.3 \spa 3.1}={\spa 1.2 {\spb 3.q}^2\over \spb 1.q \spb 2.q}, \cr
  {1\over i\sqrt{2}}V_3(1^+,2^+,3^-)&=-{{\spb 1.2}^3\over \spb 2.3
    \spb 3.1} ={\spb 2.1 {\spa 3.q}^2 \over \spa 1.q \spa 2.q}, \cr}
\equn\label{eqThreePointVertices}
$$
along with a $V_3(1^+,2^-,3^0)$ vertex.  In the formula above, all momenta are
$q$-nullified.  As vertices of this last type must be attached
together in pairs, it is natural to absorb the resulting four-point
configurations into an effective four-point vertex along with the
Yang--Mills four-point vertex.  These effective four-point vertices
contain prefactors
$$
{\spb p.q \over \spa p.q} \qquad\text{and}\qquad {\spa m.q \over \spb
  m.q}, \equn\label{eqFourPointPrefactors}
$$ 
for each positive-helicity leg $p$, and each negative-helicity leg
$m$, respectively.

When adopting a recursive approach which involves shifting a
negative-helicity leg $a$ and a positive-helicity leg $b$, the
recursion-optimised choice for the reference momentum $q$ is
\begin{equation}
\lambda _q=\lambda _a, \qquad
\bar\lambda _q=\bar\lambda _b.
\end{equation}

With this choice of $q$ the
prefactors of four-point vertices~\eqref{eqFourPointPrefactors}
involving a shifted leg vanish. 
Furthermore from \eqref{eqThreePointVertices}
the legs $a$ and $b$ can only enter a diagram on an MHV or \MHVb three-point vertex respectively.

Thus for the single-minus amplitudes, at
tree
and one-loop level, the external negative-helicity leg must enter the
diagram via an MHV three-point vertex and this must have a
negative-helicity internal leg.  This leaves insufficient negative
helicities
to have a four-point vertex anywhere in the diagram.  At tree level
there are no non-vanishing diagrams whilst at one-loop we have a
single MHV three-point vertex and several \MHVb\  three-point vertices
in each diagram.
These rules apply to both Yang--Mills and gravity calculations. For
gravity we define the tree amplitudes using the
Kawai--Lewellen--Tye (KLT) expressions~\cite{Kawai:1985xq}.

\begin{figure}[H]
  \centering\subfloat{
    \begin{picture}(120,100)
      \SetOffset(60,50)
      \Line(20,0)(40,0)
      \Line(6.1803,19.0211)(12.3607,38.0423)
      \Line(-16.1803,11.7557)(-32.3607,23.5114)
      \Line(-16.1803,-11.7557)(-32.3607,-23.5114)
      \Line(6.1803,-19.0211)(12.3607,-38.0423)
      \Vertex(20,0){1.5}
      \Vertex(6.1803,19.0211){1.5}
      \Vertex(-16.1803,11.7557){1.5}
      \Vertex(-16.1803,-11.7557){1.5}
      \Vertex(6.1803,-19.0211){1.5}
      \Line(20,0)(6.1803,19.0211)
      \Line(6.1803,19.0211)(-16.1803,11.7557)
      \Line(-16.1803,11.7557)(-16.1803,-11.7557)
      \Line(-16.1803,-11.7557)(6.1803,-19.0211)
      \Line(6.1803,-19.0211)(20,0)
      \Text(-33,25)[br]{$-$}
      \Text(12.3,40)[cb]{$+$}
      \Text(42,0)[cl]{$+$}
      \Text(12.3,-40)[bt]{$+$}
      \Text(-33,-25)[tr]{$+$}
      \Text(18,6)[cl]{\footnotesize{$+$}}
      \Text(18,-6)[cl]{\footnotesize{$-$}}
      \Text(10,17)[cl]{\footnotesize{$-$}}
      \Text(10,-17)[cl]{\footnotesize{$+$}}
      \Text(-0.5,18.5)[bc]{\footnotesize{$+$}}
      \Text(-15,14)[bl]{\footnotesize{$-$}}
      \Text(-17,10)[tr]{\footnotesize{$+$}}
      \Text(-0.5,-18.5)[tc]{\footnotesize{$-$}}
      \Text(-15,-14)[tl]{\footnotesize{$+$}}
      \Text(-17,-10)[br]{\footnotesize{$-$}}
    \end{picture}
    \label{fig:ax-pent}
  }\quad\subfloat{
    \begin{picture}(120,100)
      \SetOffset(60,50)
      \Line(-20,0)(-40,0)
      \Line(20,0)(40,0)
      \Line(0,20)(0,40)
      \Line(0,-20)(0,-40)
      \Line(40,0)(55,17.3205)
      \Line(40,0)(55,-17.3205)
      \Vertex(20,0){1.5}
      \Vertex(40,0){1.5}
      \Vertex(-20,0){1.5}
      \Vertex(0,20){1.5}
      \Vertex(0,-20){1.5}
      \Line(20,0)(0,20)
      \Line(0,20)(-20,0)
      \Line(-20,0)(0,-20)
      \Line(0,-20)(20,0)
      \Text(0,-43)[tc]{$+$}
      \Text(-43,0)[cr]{$-$}
      \Text(0,43)[bc]{$+$}
      \Text(55,20)[bc]{$+$}
      \Text(55,-20)[tc]{$+$}
      \Text(22,-1)[tl]{\footnotesize{$+$}}
      \Text(38,-1)[tr]{\footnotesize{$-$}}
      \Text(18,8)[cc]{\footnotesize{$+$}}
      \Text(8,18)[cc]{\footnotesize{$-$}}
      \Text(-18,8)[cc]{\footnotesize{$-$}}
      \Text(-8,18)[cc]{\footnotesize{$+$}}
      \Text(-18,-8)[cc]{\footnotesize{$+$}}
      \Text(-8,-18)[cc]{\footnotesize{$-$}}
      \Text(18,-8)[cc]{\footnotesize{$-$}}
      \Text(8,-18)[cc]{\footnotesize{$+$}}
    \end{picture}
    \label{fig:ax-box}
  }
  \caption{ With the constraints that (1) the negative-helicity leg
    enters via an MHV three-point vertex and (2) the four-point vertices vanish, we
    only have non-vanishing diagrams with a single three-point MHV vertex with the
    remaining vertices three-point \MHVb\ with internal helicities organised as
    shown in these sample diagrams.  }
  \label{fig:ax-oneloop}
\end{figure}
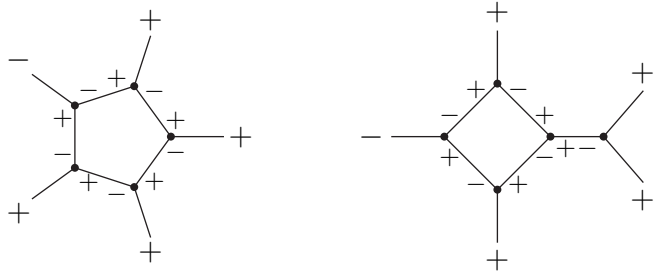

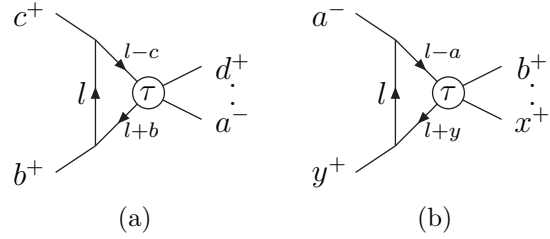
\begin{figure}[H]
  \begin{center}
    \subfloat[]{
      \begin{picture}(90,72)
        \SetOffset(12,-3)
        \ArrowLine(20,20)(20,60)
        \ArrowLine(20,60)(40,40) \ArrowLine(40,40)(20,20)
        \Line(5,10)(20,20) \Line(5,70)(20,60)
        \Line(40,40)(60,50) \Line(40,40)(60,30)
        \Text(31,24)[l]{${}^{l+{ b}}$}
        \Text(31,54)[l]{${}^{l-c}$}
        \Text(15,40)[c]{$l$}
        \Text(-5,70)[c]{$c^+$}
        \Text(-5,10)[c]{${b}^+$}
        \Text(72,50)[c]{$d^+$}
        \Text(72,43)[c]{$.$}
        \Text(72,36)[c]{$.$}
        \Text(72,30)[c]{${a}^-$}
        \BCirc(40,40){6}
        \Text(40,40)[cc]{$\Tau$}
      \end{picture}
      \label{fig:axialloop-i}
    }\quad\subfloat[]{
      \begin{picture}(90,72)
        \SetOffset(12,-3) \ArrowLine(20,20)(20,60)
        \ArrowLine(20,60)(40,40) \ArrowLine(40,40)(20,20)
        \Line(5,10)(20,20) \Line(5,70)(20,60)
        \Line(40,40)(60,50) \Line(40,40)(60,30)
        \Text(31,24)[l]{${}^{l+y}$}
        \Text(31,54)[l]{${}^{l-{a}}$}
        \Text(15,40)[c]{$l$}
        \Text(-5,70)[c]{${a}^-$}
        \Text(-5,10)[c]{$y^+$}
        \Text(72,50)[c]{${b}^+$}
        \Text(72,43)[c]{$.$}
        \Text(72,36)[c]{$.$}
        \Text(72,30)[c]{${x}^+$}
        \BCirc(40,40){6}
        \Text(40,40)[cc]{$\Tau$}
      \end{picture}
      \label{fig:axialloop-j}
    }
  \end{center}
  \caption{Singularities in $s_{bc}$ and $s_{ay}$ arise in
    integrations over the terms shown.}
  \label{fig:axialloop}
\end{figure}

We now wish to characterise the singularities when either $s_{bc}$ or
$s_{ay}$ vanish.
Singularities arise in the integration from the region of loop
momentum where the denominators of three adjacent propagators vanish
simultaneously, as the two null legs to which they connect become
collinear.
The diagrams of interest for any single-minus amplitude can then be
collected into the forms shown in \figref{fig:axialloop}. Note that we evaluate these diagrams for real momenta and only carry out analytic shifts 
on the final expressions. The
circles in these diagrams represent the sums of all possible tree
diagrams with two internal legs and the given external legs.  We denote
these by
$\Tau(a,b,\ldots)$. In the integration region of interest all the legs of
$\Tau$ are close to null and the internal legs are close to collinear.
\newcommand{\RLH}{\text{r.l.h.}}

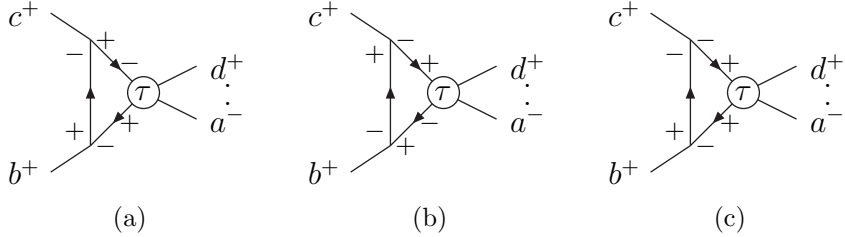
\begin{figure}[H]
  \begin{center}
    \subfloat[]{
      \begin{picture}(90,72)
        \SetOffset(12,-3)
        \ArrowLine(20,20)(20,60)
        \ArrowLine(20,60)(40,40) \ArrowLine(40,40)(20,20)
        \Line(5,10)(20,20) \Line(5,70)(20,60)
        \Line(40,40)(60,50) \Line(40,40)(60,30)
        \Text(-5,70)[c]{$c^+$}
        \Text(-5,10)[c]{${b}^+$}
        \Text(72,50)[c]{$d^+$}
        \Text(72,43)[c]{$.$}
        \Text(72,36)[c]{$.$}
        \Text(72,30)[c]{${a}^-$}
        \BCirc(40,40){6}
        \Text(40,40)[cc]{$\Tau$}
        \Text(18,58)[tr]{\footnotesize{$-$}}
        \Text(18,22)[br]{\footnotesize{$+$}}
        \Text(22,20)[cl]{\footnotesize{$-$}}
        \Text(22,60)[cl]{\footnotesize{$+$}}
        \Text(35,48)[bc]{\footnotesize{$-$}}
        \Text(35,32)[tc]{\footnotesize{$+$}}
      \end{picture}
      \label{fig:axialloop-ia}
    }\quad \subfloat[]{
      \begin{picture}(90,72)
        \SetOffset(12,-3)
        \ArrowLine(20,20)(20,60)
        \ArrowLine(20,60)(40,40) \ArrowLine(40,40)(20,20)
        \Line(5,10)(20,20) \Line(5,70)(20,60)
        \Line(40,40)(60,50) \Line(40,40)(60,30)
        \Text(-5,70)[c]{$c^+$}
        \Text(-5,10)[c]{${b}^+$}
        \Text(72,50)[c]{$d^+$}
        \Text(72,43)[c]{$.$}
        \Text(72,36)[c]{$.$}
        \Text(72,30)[c]{${a}^-$}
        \BCirc(40,40){6}
        \Text(40,40)[cc]{$\Tau$}
        \Text(18,58)[tr]{\footnotesize{$+$}}
        \Text(18,22)[br]{\footnotesize{$-$}}
        \Text(22,20)[cl]{\footnotesize{$+$}}
        \Text(22,60)[cl]{\footnotesize{$-$}}
        \Text(35,48)[bc]{\footnotesize{$+$}}
        \Text(35,32)[tc]{\footnotesize{$-$}}
      \end{picture}
      \label{fig:axialloop-ib}
    } \quad \subfloat[]{
      \begin{picture}(90,72)
        \SetOffset(12,-3)
        \ArrowLine(20,20)(20,60)
        \ArrowLine(20,60)(40,40) \ArrowLine(40,40)(20,20)
        \Line(5,10)(20,20) \Line(5,70)(20,60)
        \Line(40,40)(60,50) \Line(40,40)(60,30)
        \Text(-5,70)[c]{$c^+$}
        \Text(-5,10)[c]{${b}^+$}
        \Text(72,50)[c]{$d^+$}
        \Text(72,43)[c]{$.$}
        \Text(72,36)[c]{$.$}
        \Text(72,30)[c]{${a}^-$}
        \BCirc(40,40){6}
        \Text(40,40)[cc]{$\Tau$}
        \Text(18,58)[tr]{\footnotesize{$-$}}
        \Text(18,22)[br]{\footnotesize{$+$}}
        \Text(22,20)[cl]{\footnotesize{$-$}}
        \Text(22,60)[cl]{\footnotesize{$-$}}
        \Text(35,48)[bc]{\footnotesize{$+$}}
        \Text(35,32)[tc]{\footnotesize{$+$}}
      \end{picture}
      \label{fig:axialloop-ic}
    }
  \end{center}
  \caption{The three possible helicity structures of
    figure 4a.}
  \label{fig:axialloop-abc}
\end{figure}

In each case there are three options for the helicities within the
loop, as illustrated in \figref{fig:axialloop-abc}.
Let us consider \figref{fig:axialloop-i}.
 With the
configuration of \figref{fig:axialloop-ic}, $\tau$ vanishes at the
integration singularity because it is a one-minus tree amplitude in
this limit and so the diagram has vanishing residue.
The diagram~\ref{fig:axialloop-ib} evaluates to
$$
\int d^4l \,{ [b|l|a\ra [c|l|a\ra \over \spa{b}.{a}\spa{c}.a } \times {
  \spa{l-c,}.{a}^2 \over \spa{l+b,}.{a}^2} { \tau((l-c)^+,d^+,\cdots,
  a^-,(l+b)^-) \over l^2(l+k_b)^2(l-k_c)^2} \equn
$$ 
where the momenta in the spinor products are $q$-nullified as in
\eqref{eqqnullification}.
We construct a basis for the loop momentum built on $b$ and $c$ via
$$
l=\alpha_1(k_b+k_{c})+\alpha_2(k_b -k_{c}) +({\alpha_3}+{i\alpha_4}){
\spa{c}.a \over \spa{b}.a} \lambda_b\bar\lambda_{c}+
({\alpha_3}-{i\alpha_4})
{ \spa{b}.a \over \spa{c}.a }
\lambda _{c}\bar\lambda _b
\equn
$$
Under this parametrisation, 
$$
\int { d^4l \over l^2(l+k_b)^2(l-k_c)^2} f(l)= 
{1 \over
  s_{bc} } \int d\alpha_i \;
F(\alpha_i)f(l(\alpha_i))
\equn\label{eqParaInt}
$$
where $F(\alpha_i)$ has no dependence on  $s_{bc}$
\footnote{
We have not explicitly introduced a regulator, although even for the finite amplitudes under consideration individual diagrams may diverge. If we were to use a Pauli-Villars regulator with mass $M_{\rm PV}$ for instance, we could still extract the same momentum-dependent prefactor but the remaining integral would depend on the $\alpha_i$ and $M^2_{\rm PV}/s_{bc}$. Knowing that any divergent pieces cancel in the full amplitude allows us to consider only the finite pieces, which are independent of
$M_{\rm PV}$.}.

Also,
$$
\eqalign{
[b|l|a\ra 
&= \bigl(\alpha_1-\alpha_2+{{\alpha_3+i\alpha_4}}\bigr)[b|c|a\ra,
\cr
[c|l|a\ra 
&= \bigl(\alpha_1+\alpha_2+{{\alpha_3-i\alpha_4}}\bigr)[c|b|a\ra.
\cr}
\equn
$$
After these manipulations the integrand from \figref{fig:axialloop-i}
becomes
$$
{[bc]\over \la {b}c\ra}
  {\spa{l-c,}.{a}^2 \over \spa{l+b,}.{a}^2}
\Tau((l-c)^+,d^+,\dots,a^-,(l+b)^-)  \times F'(\alpha_i).
\equn\label{triblobYM}
$$
When $l$, $b$ and $c$ become collinear, $\Tau$ approaches the
collinear limit of an MHV tree amplitude.  Within $\Tau$ there are
diagrams with an explicit $s_{bc}$ pole.  The singular factor from the
integration around the collinear limit combines with the explicit pole
factor to give the double pole discussed previously.  In addition to
the leading behaviour of $\Tau$ in the collinear limit, we need to know
its finite piece in order to determine the residue at this pole.  The
diagram~\ref{fig:axialloop-ia} gives the same contribution.

We can apply a similar analysis to the contributions from diagrams of
the type shown in \figref{fig:axialloop-j}. In this case $\Tau$ approaches either a
one-minus or an all-plus  tree amplitude in the region of interest
and so vanishes. Thus diagrams of the type shown in  \figref{fig:axialloop-j}
give no contribution.

In order to evaluate the contribution from \eqref{triblobYM} we must
evaluate the tree structures to order $\spa{b}.c^0$. For diagrams
within $\tau$ involving $1/s_{bc}$ this means going beyond leading
order. The loop part of these diagrams is a triangle and the
calculation is readily done exactly. The diagrams without this
propagator need only be calculated to leading order.  In this regard,
not only is the recursive approach selecting a subset of diagrams for
calculation, it is also allowing us to calculate these diagrams in a
very convenient limit.

In the following section we apply an augmented recursive analysis to
the calculation of the amplitude $M^\oneloop(1^-,2^+,3^+,4^+,5^+)$.
For gravity, using the KLT relations for tree amplitudes~\cite{Kawai:1985xq}
the equivalent expression to \eqref{triblobYM} is
$$
{[bc]^3\over \la {b}c\ra}{\spa{l-c,}.{a}^4 \over \spa{l+b,}.{a}^4}
\Tau_{\text{grav}}((l-c)^+,d^+,\dots,a^-,(l+b)^-)  \times F'(\alpha_i).
\equn\label{triblob}
$$

\section{The graviton scattering amplitude
  $M^\oneloop(1^-,2^+,3^+,4^+,5^+)$}
\label{sec:grav-mpppp}
To compute this amplitude recursively, as discussed in the previous
section, we must compute three types of contribution:
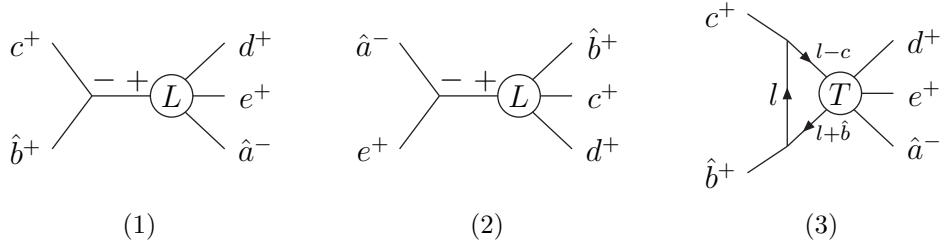
\[\begin{matrix}
  \begin{picture}(100,72)
    \SetOffset(22,-4)
    \CArc(40,40)(8,0,360)
    \Line(-5,20)(10,40) \Line(-5,60)(10,40)
    \Line(10,40)(32,40)
    \Line(45,46)(60,60) \Line(45,34)(60,20) \Line(48,40)(60,40)
    \Text(-15,60)[c]{$c^+$} \Text(40,40)[c]{$L$} \Text(-15,20)[c]{${\hat
        b}^+$} \Text(72,60)[c]{$d^+$} \Text(72,40)[c]{$e^+$}
    \Text(72,20)[c]{${\hat a}^-$}
    \Text(10,42)[bl]{$-$}
    \Text(32,42)[br]{$+$}
  \end{picture}&\qquad&
  \begin{picture}(100,72)
    \SetOffset(22,-4)
    \CArc(40,40)(8,0,360)
    \Line(-5,20)(10,40) \Line(-5,60)(10,40) \Line(10,40)(32,40)
    \Line(45,46)(60,60) \Line(45,34)(60,20) \Line(48,40)(60,40)
    \Text(-15,60)[c]{$\hat a^-$} \Text(40,40)[c]{$L$}
    \Text(-15,20)[c]{${e}^+$} \Text(72,60)[c]{${\hat b}^+$}
    \Text(72,40)[c]{$c^+$} \Text(72,20)[c]{${d}^+$}
    \Text(10,42)[bl]{$-$}
    \Text(32,42)[br]{$+$}
  \end{picture}&\qquad&
  \begin{picture}(90,72)
    \SetOffset(12,-3)
    \ArrowLine(20,20)(20,60)
    \ArrowLine(20,60)(35,46) \ArrowLine(35,34)(20,20)
    \CArc(40,40)(8,0,360) \Line(5,10)(20,20) \Line(5,70)(20,60)
    \Line(45,46)(60,60) \Line(45,34)(60,20) \Line(48,40)(60,40)
    \Text(40,40)[c]{$T$} \Text(31,24)[l]{${}^{l+{\hat b}}$}
    \Text(31,54)[l]{${}^{l-c}$} \Text(15,40)[c]{$l$}
    \Text(-5,70)[c]{$c^+$} \Text(-5,10)[c]{${\hat b}^+$}
    \Text(72,60)[c]{$d^+$} \Text(72,40)[c]{$e^+$}
    \Text(72,20)[c]{${\hat a}^-$}
  \end{picture}\\
  \mcap{(1)} & & \mcap{(2)} & & \mcap{(3)}   
\end{matrix}\]
together with the contributions
obtained by summing over the distinct permutations of $c$, $d$ and
$e$.

The first two of these involve single poles only, so we only
need the loop structures to leading order and we can use the
corresponding four-pt one-loop amplitudes. The final structure contains a
double pole so we must evaluate both the tree structure on the right
and the loop pieces more carefully.  The first diagram uses the
four-point one-minus amplitude whereas the second requires the
four-point all-plus amplitude, both are given in \eqref{eqFourPoints}.
We obtain,
$$
\eqalign{ \Rec_1(a,b,c,d,e)&={1 \over 5760}{\la ad\ra^2 \la ae\ra^2
[bc][de]^4
    \left( \la cd\ra^2 \la ae\ra^2+\la ac\ra\la cd\ra\la de\ra\la
      ae\ra+\la ac\ra^2\la de\ra^2 \right)\over \la ab\ra^2\la
    bc\ra\la ce\ra^2\la cd\ra^2\la de\ra^2}, \cr
  \Rec_2(a,b,c,d,e)&=-{3\over 5760}{\la ae\ra [be]^4 \over \la cd\ra^2
[ab]^2
    [ae]}\left([bc]^2[ de]^2+[ bc][ cd][ de][ be]+[ cd]^2[
    be]^2\right).  \cr}\equn\label{eq:R1and2}
$$

For the third diagram, we need the tree diagrams which constitute
$\Tau_{\text{grav}}$ of \eqref{triblob}. 
Mindful of the recursive analysis that we will ultimately perform, we
calculate
these diagrams as Laurent series in $\spa{b}.c$, dropping terms that will
not contribute to the residues.

 We require the five-point
contributions with two off-shell legs $B^-$ and $C^+$,
carrying momenta
 $B\equiv l+b$ and $C\equiv
c-l$, respectively.

\def\COMMENT{ 
We use the KLT relations~\cite{Kawai:1985xq} for five points to
obtain this in terms of Yang--Mills amplitudes,
$$
\eqalign{ M(a^-,B^-,C^+,d^+,e^+) =& s_{BC}s_{de}
  A(a^-,B^-,C^+,d^+,e^+) A(a^-,C^+,B^-,e^+,d^+) \cr +& s_{Bd} s_{Ce}
  A(a^-,B^-,d^+,C^+,e^+)A(a^-,d^+,B^-,e^+,C^+), \cr}
\equn\label{eqKLT5pt}
$$
where we have chosen a form of the KLT relations that restricts the
$\la bc\ra$ pole to the first term.  

Strictly, the KLT relations are only valid for on-shell momenta, however these momenta may be in dimensions higher than four.  
A study of the more general situation 

Strictly, the KLT relations are only valid for on-shell momenta, however these momenta may be in dimensions higher than four.  
We find this allows us to compute the singularity as
$$
\la bc\ra\biggl({T^{\rm leading}\over \la bc\ra}+T^{\text{sub-leading}}\biggr)\biggl({T^{\rm leading}\over \la bc\ra}+T^{\text{sub-leading}}\biggr),
\equn
$$
where all diagrams contribute to the sub-leading pieces but only diagrams involving a $V_3(B^-,C^+,x)$ vertex contribute to the leading pieces.
The second term in eq.~\ref{eqKLT5pt} is only
needed to leading order and its contribution to the residue will be directly determined by the
on-shell \YM MHV amplitudes.

\COMMENT}

The KLT relation~\cite{Kawai:1985xq} 
between Yang--Mills amplitudes and gravity amplitudes at five points is
$$
\eqalign{ M(a^-,B^-,C^+,d^+,e^+) =& s_{BC}s_{de}
  A(a^-,B^-,C^+,d^+,e^+) A(a^-,C^+,B^-,e^+,d^+) \cr +& s_{Bd} s_{Ce}
  A(a^-,B^-,d^+,C^+,e^+)A(a^-,d^+,B^-,e^+,C^+), \cr}
\equn\label{eqKLT5pt}
$$
where we have chosen a form of the KLT relations that restricts the
$\la bc\ra$ pole to the first term.  
The KLT relations are only valid for on-shell momenta, although these momenta may be in higher dimensions.
If we assume the deviation from eq.~(\ref{eqKLT5pt}) may be neglected in the region around $B^2=C^2=0$,
we see that the gravity tree structure has the form,
$$
\la bc\ra\biggl({T^{\rm leading}\over \la bc\ra}+T^{\text{sub-leading}}\biggr)\biggl({T^{\rm leading}\over \la bc\ra}+T^{\text{sub-leading}}\biggr),
\equn
$$
where all diagrams contribute to the sub-leading pieces but only diagrams involving a $V_3(B^-,C^+,x)$ vertex contribute to the leading pieces.
The second term in eq.~(\ref{eqKLT5pt}) is only
needed to leading order and its contribution to the residue will be directly determined by the
on-shell \YM MHV amplitudes. The amplitude generated using the leading and sub-leading singularity terms from (\ref{eqKLT5pt}) has 
the correct symmetries and collinear limits.  Additionally, the five-point amplitude has been verified by 
a completely independent string-based rules computation.  
The general case is worthy of further study~\cite{Alstonetal}.

First we establish the double pole term. This arises from the poles in
each of the Yang--Mills tree amplitudes in the first term of
\eqref{eqKLT5pt}.  We evaluate this diagrammatically. The \YM
amplitude, $A(a^-,B^-,C^+,d^+,e^+)$ receives contributions from five
diagrams. The two which contribute to the pole are:
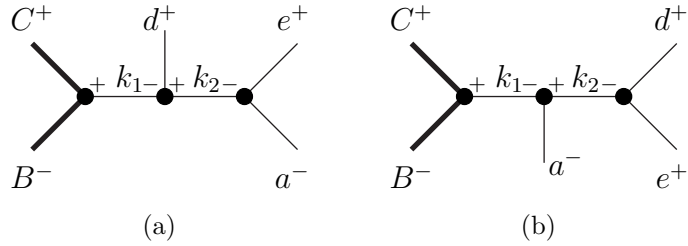
\[\begin{matrix}
  \begin{picture}(112,72)
    \SetOffset(8,-4)
    \SetWidth{2.0} \Line(0,20)(20,40)
    \Line(0,60)(20,40) \SetWidth{0.5} \Line(20,40)(80,40)
    \Line(80,40)(100,20) \Line(80,40)(100,60) \Line(50,40)(50,65)
    \CCirc(20,40){3}{0}{0} \CCirc(50,40){3}{0}{0}
    \CCirc(80,40){3}{0}{0} \Text(37,47)[c]{${k_1}$}
    \Text(66,47)[c]{${k_2}$} \Text(25,43)[c]{${}^+$}
    \Text(45,43)[c]{${}^-$} \Text(55,43)[c]{${}^+$}
    \Text(75,43)[c]{${}^-$} \Text(0,70)[c]{$C^+$}
    \Text(0,10)[c]{$B^-$} \Text(55,70)[r]{$d^+$}
    \Text(105,70)[r]{$e^+$} \Text(105,10)[r]{$a^-$}
  \end{picture}
  &\qquad&
  \begin{picture}(112,72)
    \SetOffset(8,-4)
    \SetWidth{2.0} \Line(0,20)(20,40)
    \Line(0,60)(20,40) \SetWidth{0.5} \Line(20,40)(80,40)
    \Line(80,40)(100,20) \Line(80,40)(100,60) \Line(50,40)(50,15)
    \CCirc(20,40){3}{0}{0} \CCirc(50,40){3}{0}{0}
    \CCirc(80,40){3}{0}{0} \Text(37,47)[c]{${k_1}$}
    \Text(66,47)[c]{${k_2}$} \Text(25,43)[c]{${}^+$}
    \Text(45,43)[c]{${}^-$} \Text(55,43)[c]{${}^+$}
    \Text(75,43)[c]{${}^-$} \Text(0,70)[c]{$C^+$}
    \Text(0,10)[c]{$B^-$} \Text(105,70)[r]{$d^+$}
    \Text(105,10)[r]{$e^+$} \Text(65,15)[r]{$a^-$}
  \end{picture} \\
  \mcap{(a)} & & \mcap{(b)}
\end{matrix}\] with contributions
$$
\eqalign{ D_a &= {\la Ba\ra^2\over \la Ca\ra^2}{\la a|bc|a\ra [de]
    [eb]\over s_{bc} \la da\ra \la ea\ra [ae][ab]} f_a(\alpha_i) , \cr
  D_b &= {\la Ba\ra^2\over\la Ca\ra^2}{\la a|bc|a\ra \la ca\ra \over
    \la bc\ra [ab]\la de\ra\la da\ra\la ea\ra} f_b(\alpha_i),
  \cr}\equn
$$
where we have used $\la a|BC|a\ra=\la a|(b+l)(c-l)|a\ra = \la
a|bc|a\ra f(\alpha_i)$, \etc.  The parameters contained in $f_a$ and
$f_b$ are the same for both diagrams and the sum of the two
contributions is
\begin{equation}
  D_a+D_b=
  {\la Ba\ra^2\over\la Ca\ra^2} {\la a|bc|a\ra \over
    s_{bc} [ab]\la da\ra\la ea\ra}
  \Biggl( 
  {[b|ad|e]-[b|cb|e]\over [ae]\la  de\ra}
  \Biggr) f_a(\alpha_i),
  \label{BCfuse}
\end{equation}
where the second term is sub-leading in the $\la bc\ra$ pole.

The leading pole in the other Yang--Mills factor is obtained
analogously and, combining, we obtain the leading pole in
\eqref{eqKLT5pt},
$$
{\la Ba\ra^4\over\la Ca\ra^4} s_{bc}s_{de}{\la ab\ra \la ac\ra [de]
  \over \la bc\ra \la ea\ra[ ae]\la de\ra} {\la ab\ra \la ac\ra [de]
  \over \la bc\ra \la da\ra[ ad]\la de\ra}
 f_a'(\alpha_i)  . \equn
$$
Combining this with the factors arising from the left hand part of the full
diagram
and integrating over the $\alpha_i$
the leading term in the Laurent series is proportional to
$$
{[bc]^3\over \la bc\ra}[bc][de]{\la ab\ra \la ac\ra [de]
  \over \la ea\ra[ ae]\la de\ra} {\la ab\ra \la ac\ra [de]
  \over \la bc\ra \la da\ra[ ad]}, \equn
\label{eq:Rec3}
$$
which clearly displays the double pole factor. The constant of proportionality is most readily fixed by looking at collinear limits. 

We must now enumerate the contributions that are sub-leading in the $s_{bc}$
pole.  These come from a variety of sources.  We express these
single-pole terms as the double-pole factor of \eqref{eq:Rec3},
multiplied by a factor $\delta$.  Firstly, we have the sub-leading
contribution of \eqref{BCfuse} together with the corresponding contribution from the other Yang-Mills factor,
$$
\delta_1 = {s_{bc}[be]\over [b|ad|e]}+ {s_{bc}[bd]\over [b|ae|d]}.
\equn
$$
Next we have the sub-leading diagrams for the Yang--Mills amplitudes
in the first term of \eqref{eqKLT5pt} shown below:
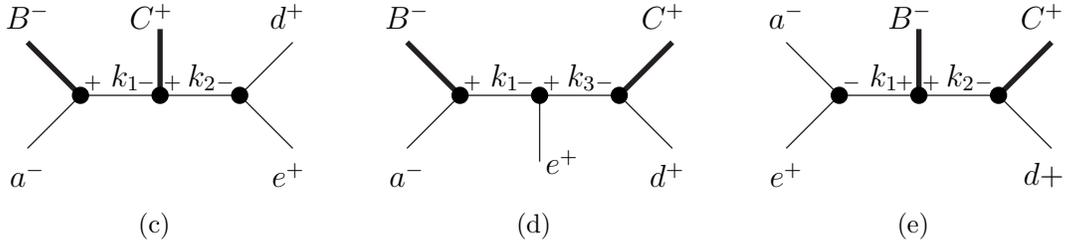
\[\begin{matrix}
  \begin{picture}(112,72)
    \SetOffset(8,-4)
    \SetWidth{2.0} \Line(50,40)(50,65)
    \Line(0,60)(20,40) \SetWidth{0.5} \Line(0,20)(20,40)
    \Line(20,40)(80,40) \Line(80,40)(100,20) \Line(80,40)(100,60)
    \CCirc(20,40){3}{0}{0} \CCirc(50,40){3}{0}{0}
    \CCirc(80,40){3}{0}{0} \Text(37,47)[c]{${k_1}$}
    \Text(66,47)[c]{${k_2}$} \Text(25,43)[c]{${}^+$}
    \Text(45,43)[c]{${}^-$} \Text(55,43)[c]{${}^+$}
    \Text(75,43)[c]{${}^-$} \Text(0,70)[c]{$B^-$}
    \Text(0,10)[c]{$a^-$} \Text(55,70)[r]{$C^+$}
    \Text(105,70)[r]{$d^+$} \Text(105,10)[r]{$e^+$}
  \end{picture}
  &\qquad&
  \begin{picture}(112,72)
    \SetOffset(8,-4)
    \SetWidth{2.0} \Line(80,40)(100,60)
    \Line(0,60)(20,40) \SetWidth{0.5} \Line(0,20)(20,40)
    \Line(20,40)(80,40) \Line(80,40)(100,20) \Line(50,40)(50,15)
    \CCirc(20,40){3}{0}{0} \CCirc(50,40){3}{0}{0}
    \CCirc(80,40){3}{0}{0} \Text(37,47)[c]{${k_1}$}
    \Text(66,47)[c]{${k_3}$} \Text(25,43)[c]{${}^+$}
    \Text(45,43)[c]{${}^-$} \Text(55,43)[c]{${}^+$}
    \Text(75,43)[c]{${}^-$} \Text(0,70)[c]{$B^-$}
    \Text(0,10)[c]{$a^-$} \Text(105,70)[r]{$C^+$}
    \Text(105,10)[r]{$d^+$} \Text(65,15)[r]{$e^+$}
  \end{picture}
  &\qquad&
  \begin{picture}(112,72)
    \SetOffset(8,-4)
    \SetWidth{2.0} \Line(50,40)(50,65)
    \Line(80,40)(100,60) \SetWidth{0.5} \Line(0,20)(20,40)
    \Line(0,60)(20,40) \Line(20,40)(80,40) \Line(80,40)(100,20)
    \CCirc(20,40){3}{0}{0} \CCirc(50,40){3}{0}{0}
    \CCirc(80,40){3}{0}{0} \Text(37,47)[c]{${k_1}$}
    \Text(66,47)[c]{${k_2}$} \Text(25,43)[c]{${}^-$}
    \Text(45,43)[c]{${}^+$} \Text(55,43)[c]{${}^+$}
    \Text(75,43)[c]{${}^-$} \Text(0,70)[c]{$a^-$}
    \Text(0,10)[c]{$e^+$} \Text(55,70)[r]{$B^-$}
    \Text(105,70)[r]{$C^+$} \Text(105,10)[r]{$d+$}
  \end{picture} \\
  \mcap{(c)} & & \mcap{(d)} & & \mcap{(e)}
\end{matrix}\] 
We note that in the first two of these the $k_1$
propagator feeds into the two diagrams that would make a one-minus
four-point tree if $k_1$ and $C$ were both null. As we know that this
vanishes when all the legs are null, the sum of the first two diagrams
must be of the form: $C^2 X +k_1^2 Y$.  We can drop the terms
containing a $C^2$ factor as we are already at sub-leading order,
leaving something proportional $k_1^2$. Thus taking both terms
together leads to the cancellation of the $s_{aB}$ propagator:
$$
D_c+D_d= {\la Ba\ra^2\over \la Ca\ra^2} {[b|B|a\ra\over [ab] \la
  ea\ra\la de\ra} {\la ca\ra\over \la cd\ra}f_c(\alpha_i) +{\cal O}(\la bc\ra).
\equn
$$
Pulling out a factor of \eqref{eq:Rec3} leaves
$$
\delta_2={s_{bc}[e|a|c\ra \over s_{ab}[e|d|c\ra }.
\equn
$$
We can apply the same procedure to the final diagram giving
$$
\delta_3= {\la bc\ra \la de\ra \over s_{ab}[de]}\biggl( {[e|B|a\ra
  [eb] \over \la da\ra \la cd\ra}+{[d|B|a\ra [db] \over \la ea\ra \la
  ce\ra} \biggr)
\equn
$$
Finally we need the second term in \eqref{eqKLT5pt}, $s_{Bd}s_{Ce}
A(a,B,d,C,e) A(a,d,B,e,C)$, which we evaluate using
MHV tree amplitudes. After extracting the double-pole factor we
obtain
$$
\delta_4 ={\la bc\ra\la de\ra [d|B|a\ra[e|C|a\ra \over [bc][de]\la
  ab\ra^2\la cd\ra\la ce\ra}.
  \equn
$$
We thus have the leading and sub-leading poles expressed as
$$
{[bc]^3\over \la bc\ra}[bc][de]{\la ab\ra \la ac\ra [de]
  \over \la ea\ra[ ae]\la de\ra} {\la ab\ra \la ac\ra [de]
  \over \la bc\ra \la da\ra[ ad]}
  \times \left( 1 +\sum_i \delta_i \right) \equn
$$

We can now use the pole to determine the amplitude recursively. This
involves applying the shift \eqref{eqshift} and evaluating at $z=-{\la bc\ra/\la ac\ra}$. 
The coefficient of the double pole in \eqref{eq:Rec3} has a $z$
dependence under this shift which generates a further contribution to the
single pole since
$$
\Res \left( { f(z) \over z(z-z_i)^2 }, z_i \right)= -{ f(z_i)\over
  z_i^2 } + { 1 \over z_i } \left.\frac{df}{dz}\right|_{z=z_i}. 
  \equn
$$
Carrying this out and combining with the contributions of the
$\delta$s gives
\begin{equation}
  \begin{split}
    \Delta(a,b,c,d,e)&= -{1\over2}{\la ad\ra\la bc\ra\over \la
      ab\ra\la cd\ra} -{1\over2}{\la ae\ra\la bc\ra\over \la ab\ra\la
      ce\ra} \\
    & \quad -3{[db][eb]\over \la dc\ra\la ec\ra}{\la bc\ra\over
      [bc]}{\la de\ra\over [de]} -3{[dc][ec]\over \la dc\ra\la
      ec\ra}{\la bc\ra\over [bc]}
    {\la de\ra\over [de]}{\la ca\ra^2\over\la ba\ra^2}\\
    & \quad -{7\over2}{[dc][eb]\over\la dc\ra\la ec\ra}{\la
      bc\ra\over[bc]}{\la de\ra\over[de]}{\la ca\ra\over\la ba\ra}
    -{7\over2}{[db][ec]\over\la dc\ra\la ec\ra}{\la
      bc\ra\over[bc]}{\la de\ra\over[de]}{\la ca\ra\over\la ba\ra}.
  \end{split}
\end{equation}

The full one-minus amplitude can now be written as the sum over
recursive contributions arising from three orderings of the external
legs,
\begin{equation}
  M^\oneloop(1^-,2^+,3^+,4^+,5^+) = \Rec(1,2,3,4,5) + \Rec(1,2,4,5,3)
  + \Rec(1,2,5,3,4).
  \label{eq:recursion-amp}
\end{equation}
with the full amplitude having a factor of $i\kappa^5/16\pi^2$ as in
(\ref{eqOverallFactors}).

Each recursive term is a sum over the three classes of recursive
diagram,
\begin{equation}
  \Rec(a,b,c,d,e)=\Rec_1(a,b,c,d,e)+\Rec_2(a,b,c,d,e)+\Rec_3(a,b,c,d,e),
\end{equation}
where $R_1$ and $R_2$ are given by \eqref{eq:R1and2}, and
$$
\Rec_3(a,b,c,d,e)={1 \over 5760} {\la ab\ra^2\la ac\ra^4[bc]^4[de] \over\la
ad\ra\la
  ae\ra\la bc\ra^2\la cd\ra\la ce\ra\la de\ra}
\bigl(1+\Delta(a,b,c,d,e)\bigr).
\equn
$$
The overall normalisation can be obtained by evaluating the parameter
integrals or, more easily, fixed by factorising the known four-point
amplitude. 
The individual factors on the terms in $\Delta$ are also obtainable by
parameter integration or more conveniently by the
normalisation of the collinear limits.

This form for the amplitude has the correct collinear limits and is
symmetric under interchange of any pair of positive-helicity legs.  We
have also checked that the amplitude agrees with that calculated by
string-based rules.  This calculation can readily be extended to the
six-point case, $M^\oneloop(1^-,2^+,3^+,4^+,5^+,6^+)$.  We have
constructed the amplitude and again checked that it has the correct
symmetries and collinear limits.
This result is presented in
\appref{app-sixpoint}. \textit{Mathematica} code for both the five-
and six-point amplitudes may be found at
\url{http://pyweb.swan.ac.uk/~dunbar/graviton.html}.

\section{Conclusions and remarks}

In this article we have demonstrated how to augment recursion in order to determine the
rational terms in amplitudes with double poles under a complex
shift. 
Double poles are generic in amplitudes, however it is
often possible to carry out a recursion which avoids then. 
However, double poles are unavoidable in the case of the one-minus
Yang--Mills amplitudes $A^\oneloop(1^-,2^+,3^+,\ldots ,n^+)$ and the gravity
amplitudes $M^\oneloop(1^-,2^+,3^+,\ldots ,n^+)$.
In the absence of a universal soft factor analogous to
\eqref{eqYangMillsSoft}, in order to perform the augmented recursion the
sub-leading poles must be determined
on a case-by-case basis. While we have done this for both the five- and
six-point one-minus gravity amplitudes, this procedure could be used
to calculate the seven-point or indeed any higher-point one-minus
amplitude. 

\appendix

\section{Six-point single-minus amplitude}
\label{app-sixpoint}

\newcommand{\Recsix}{\Rec^{(6)}} \def\bartree{\Recsix_1}
\def\lamtree{\Recsix_2} \def\lamloop{\Recsix_3}
\def\fourfour{\Recsix_4}
\newcommand{\sumA}{\sum_{\substack{x \in \{3,4,5,6\} \\
      \{y_1,y_2,y_3\} \cup \{x\} = \{3,4,5,6\}}}}
\newcommand{\sumB}{\sum_{\substack{\{x_1,x_2\} \subset \{3,4,5,6\} \\
      \{y_1,y_2\} \cup \{x_1,x_2\} = \{3,4,5,6\} }}}
\def\finbits{{\rm KLT}_{\text{F}}} \def\YML{{\rm YM}_{\text{L}}}
\def\YMS{{\rm YM}_{\text{S}}}

The six-point one-loop single-minus graviton scattering amplitude can
also be calculated using augmented recursion. The calculation follows
that of the five-point amplitude with the addition of factorisations
involving a four-point tree amplitude and a four-point loop
amplitude. The shift employed is once again $\bar\lambda_1\to
\bar\lambda_1-z\bar\lambda_2$, $\lambda_2\to\lambda_2+z\lambda_1$. The
amplitude is given by
\begin{multline}
  M^\oneloop(1^-,2^+,3^+,4^+,5^+,6^+) = \\
  \sumA \text{ $\displaystyle \begin{aligned}[t]\bigl\{ &
      \bartree(1,x|2,y_1,y_2,y_3;z_x) +
      \lamtree(2,x|1,y_1,y_2,y_3;z_x) \\ & +
      \lamloop(2,x|1,y_1,y_2,y_3;z_x) \bigr\} \end{aligned} $ }
  \\
  + \sumB \fourfour(1,x_1,x_2|2,y_1,y_2;z_{x_1,x_2}).
\end{multline}
In each of these terms the vertical bar denotes a split of the external momenta
with the relevant pole arising when the shifted total momentum to the
right of bar is null.

The $\bartree$ terms are the factorisations involving a three-point
\MHVb\ tree and a five-point all-plus one-loop amplitude:
$$
\bartree(1,x|2,y_1,y_2,y_3;z_x)= {\spb x.2^2 \spa 1.x \over \spb 1.2^2
  \spb 1.x}M^{\oneloop}(p^+,\hat 2^+,y_1^+,y_2^+,y_3^+),
 \equn
$$
with $ z_x=\spb 1.x/\spb 2.x$ and $p=(\lambda^x+\lambda^1 \spb 1.2 /
\spb x.2) \bar\lambda^x$.  Similarly, the $\lamtree$ terms are the
factorisations involving a three-point MHV tree and a five-point
one-minus one-loop amplitude:
$$
\lamtree(2,x|1,y_1,y_2,y_3;z_x)={\spa 1.x^2 \spb 2.x \over \spa 1.2^2
  \spa 2.x}M^\oneloop(\hat 1^-,y_1^+,y_2^+,y_3^+,p^+),
\equn
$$
with $z_x=-\spa x.2/\spa x.1$ and $p=\lambda^x (\bar\lambda^x+ \spa
1.2 \bar\lambda^2/ \spa 1.x )$.  The $\fourfour$ terms are the
factorisations involving a four-point MHV tree and a four-point
all-plus one-loop amplitudes:
$$
\fourfour(1,x_1,x_2|2,y_1,y_2;z_{x_1x_2})= {M^{\tree}(\hat
  1^-,-p^-,x_1^+,x_2^+)M^{\oneloop}(\hat 2^+,y_1^+,y_2^+,p^+)\over
  t_{1x_1x_2}},
\equn$$
with $z_{x_1x_2}=t_{1x_1x_2}/[2|P_{x_1x_2}|1\ra$ and $p=k_{\hat
  1}+k_{x_1}+k_{x_2}$.

The $\lamloop$ terms are the augmented pieces arising from the $\spa 2.x$ poles. Here, $z_x=-\spa x.2/\spa x.1$.
\begin{equation}\begin{split}
    &\lamloop(2,x|1,y_1,y_2,y_3;z_x)= {[2x]^3\over \la
      2x\ra}\biggl\{-\finbits(\hat 1,\hat 2,x,y_1,y_2,y_3) \\ &
    +{[2x]\over\la 2x\ra}s_{y_2y_3} \YML(\hat 1,\hat 2,x,y_1,y_2,y_3)
    \\ &\qquad \times [s_{y_1y_3}\YML(\hat 1,\hat 2,x,y_3,y_1,y_2) +
    (s_{y_1y_2}+s_{y_1y_3})\YML(\hat 1,\hat 2,x,y_3,y_2,y_1)] \\ &
    +[2x]s_{y_2y_3} \YML(\hat 1,\hat 2,x,y_1,y_2,y_3) \\ &\qquad
    \times [s_{y_1y_3}\YMS(\hat 1,\hat 2,x,y_3,y_1,y_2) +
    (s_{y_1y_2}+s_{y_1y_3})\YMS(\hat 1,\hat 2,x,y_3,y_2,y_1)]\\ &
    +[2x]s_{y_2y_3} \YMS(\hat 1,\hat 2,x,y_1,y_2,y_3) \\ &\qquad
    \times [s_{y_1y_3}\YML(\hat 1,\hat 2,x,y_3,y_1,y_2) +
    (s_{y_1y_2}+s_{y_1y_3})\YML(\hat 1,\hat 2,x,y_3,y_2,y_1)]\\ &
    +{[2x]\over \la 2x\ra}s_{y_1y_3} \YML(\hat 1,\hat 2,x,y_2,y_1,y_3)
    \\ &\qquad \times [s_{y_2y_3}\YML(\hat 1,\hat 2,x,y_3,y_2,y_1) +
    (s_{y_2y_1}+s_{y_2y_3})\YML(\hat 1,\hat 2,x,y_3,y_1,y_2)]\\ &
    +[2x]s_{y_1y_3} \YML(\hat 1,\hat 2,x,y_2,y_1,y_3) \\ &\qquad
    \times [s_{y_2y_3}\YMS(\hat 1,\hat 2,x,y_3,y_2,y_1) +
    (s_{y_2y_1}+s_{y_2y_3})\YMS(\hat 1,\hat 2,x,y_3,y_1,y_2)]\\ &
    +[2x]s_{y_1y_3} \YMS(\hat 1,\hat 2,x,y_2,y_1,y_3) \\ &\qquad
    \times [s_{y_2y_3}\YML(\hat 1,\hat 2,x,y_3,y_2,y_1) +
    (s_{y_2y_1}+s_{y_2y_3})\YML(\hat 1,\hat 2,x,y_3,y_1,y_2)]
    \biggr\},
  \end{split}\end{equation}
where the KLT terms contributing to the double pole have leading
Yang-Mills factors:
\begin{multline}
  \YML(a,b,c,d,e,f)= {\la ac\ra \la ba \ra \over [ab]\la da\ra\la
    ea\ra\la fa\ra}\Biggl\{ {[bc]\la ca\ra \over
    t_{def}}\left({[f|P_{de}|a\ra\over \la de\ra}
    +{[d|P_{ef}|a\ra\over \la fe\ra}\right)
  \\
  +{[d|P_{ef}|a\ra[b|P_{ef}|a\ra\over \la ef\ra t_{efa}}
  +{[d|P_{ef}|a\ra[ef][fb]\over [af] t_{efa}} +{[f|P_{de}|a\ra
    [fb]\over [af]\la de\ra}\Biggr\},
\end{multline}
and sub-leading factors:
\begin{equation}\begin{split}
    \YMS(a,b,c,d,e,f)&= {\la ba \ra \over [ab]\la da\ra\la ea\ra\la
      fa\ra}\Biggl\{ {[d|P_{ef}|a\ra[b|P_{ef}|a\ra^2\over \la ef\ra
      t_{efa}^2}
    \\
    & \quad+{[d|P_{ef}|a\ra[ef][fb]\over [af] t_{efa}} \left(
      {[b|P_{ef}|a\ra\over t_{efa}} + {[bf]\over [af]}\right)
    -{[f|P_{de}|a\ra [bf]^2\over [af]^2\la de\ra}\Biggr\}
    \\
    &\quad +{1\over 2}{[bc]\la ac\ra^2\over [ab]\la cd\ra \la de\ra\la
      ef\ra\la fa\ra} -{1\over 2}{[f|k_b-k_c|a\ra \la ac\ra[fb]\over
      [ab]\la cd\ra \la de\ra\la ea\ra s_{fa}}
    \\
    &\quad -{1\over 2}{\la a|(k_b-k_c)P_{ef}|a\ra \la ac\ra \over
      [ab]\la cd\ra \la da\ra\la ea\ra t_{efa}} \left({
        [b|P_{ef}|a\ra \over \la fe\ra \la fa\ra} +{[bf][ef]\over
        s_{fa}}\right).
  \end{split}\end{equation}
Finally the finite terms in the KLT sum are:
\begin{equation}\begin{split}
    &\finbits(a,b,c,d,e,f)=
    \\
    &\quad {s_{ef}\la ab\ra^4\over 2\la ad\ra\la ae\ra \la af\ra \la
      bd\ra^2\la
      be\ra^2\la bf\ra\la ef\ra^2} \\
    &\quad \times \biggl\{ 
     \la ae\ra 
     \bigl( 6 \la a|bP_{ef}|b\ra [d|b|a\ra +7 \la a|bP_{ef}|b\ra [d|c|a\ra +7 \la a|cP_{ef}|b\ra [d|b|a\ra +6 \la a|cP_{ef}|b\ra [d|c|a\ra \bigr)\\
     &\qquad 
     +\la ab\ra 
     \bigl( 6 \la e|fb|a\ra [d|b|a\ra +7 \la e|fb|a\ra [d|c|a\ra +7 \la e|fc|a\ra [d|b|a\ra +6 \la e|fc|a\ra [d|c|a\ra \bigr) \biggr\}
    \\
    & +{s_{df}\la ab\ra^4\over 2\la ad\ra\la ae\ra \la
      bd\ra\la be\ra\la bf\ra^2\la de\ra\la df\ra}
    \\
    & \quad \times\bigl\{ 6 [e|b|a\ra [f|b|a\ra +7 [e|b|a\ra [f|c|a\ra
    +7 [e|c|a\ra [f|b|a\ra +6 [e|c|a\ra [f|c|a\ra \bigr\} \cr & +{\la
      ab\ra^4\over 2\la ad\ra\la ae\ra \la af\ra \la bd\ra\la be\ra\la
      bf\ra^2\la de\ra} \bigl\{
    [d|c|a\ra[e|b|a\ra[f|c|a\ra+[d|b|a\ra[e|c|a\ra[f|b|a\ra\bigr\}
    \\
    & +\{d\leftrightarrow e\}
  \end{split}\end{equation}

Expressed na\"ively, without attempting optimisation, as a rational polynomial 
of the $\lambda_{\alpha}^i$ it has a {\tt LeafCount} 
of 355,053.  For 
comparison, the {\tt LeafCount} of the five-point one-minus gravity amplitude is
4,549, and for the six-point one-minus Yang--Mills amplitude is 1,541.

\section{Graviton scattering amplitudes}
\label{sec:gravscat}
We define tree and
one-loop amplitudes in gravity for which all field couplings have been
removed, {\it i.e.},
$$
\eqalign{
  {\cal M}_n^\tree(1,2,\ldots, n) &= 
i \kappa^{(n-2)} M_n^\tree(1,2,\ldots, n),\cr
  {\cal M}_n^\oneloop(1,2,\ldots, n) &= {i\kappa^n \over (4\pi)^2 }
   M^\oneloop_n (1,2,\ldots, n ).
\cr}
\equn\label{eqOverallFactors}$$
As for Yang--Mills amplitudes we express amplitudes using the spinor
helicity formalism. For the four dimensional case there are only two graviton
helicities and their polarisation tensors can be constructed from direct
products of Yang--Mills polarisations vectors,
\begin{equation}
  \label{eq:gravpol}
  \varepsilon_{\mu\nu}^{+} = \varepsilon_{\mu}^+ \, \varepsilon_{\nu}^+,
  \qquad \varepsilon_{\mu\nu}^{-}  = \varepsilon_{\mu}^- \,
  \varepsilon_{\nu}^-.
\end{equation}

If we consider the Feynman diagrams for a gravity one-loop scattering
amplitude, performing a Passarino--Veltman reduction~\cite{Reduction} allows us to
reduce any one-loop amplitude to the form
$$
M^{\oneloop}_n(1,\dots,n)=\sum_{i}\, c_i\, I_4^{i}
+\sum_{j}\, d_{j}\, I_3^{j} +\sum_{k}\, e_{k} \,
I_2^{k} +R\, +O(\epsilon).
\equn
$$

Relatively few graviton scattering amplitudes have been computed.  In
fact, only the four-point amplitudes have been computed for all
helicity configurations~\cite{Bern:1993wt, Dunbar:1994bn,Grisaru:1979re,Dunbar:2002gu,Dunbar:1999nj} and 
all possible matter types circulating in the loop.
For four points there are three independent helicity configurations
for the external gravitons: 
$M(1^+,2^+,3^+,4^+)$, $M(1^-,2^+,3^+,4^+)$ and  $M(1^-,2^-,3^+,4^+)$.
The
all-plus and one-minus vanish at tree level and have one-loop
amplitudes which are purely rational (to order $\epsilon^0$). These
amplitudes, for any matter content, are
$$
\eqalign{
  M^\oneloop(1^+,2^+,3^+,4^+) &= - N_s  \Bigl({s t
    \over \spa1.2\spa2.3\spa3.4\spa4.1} \Bigr)^2 {( s^2 +st +t^2 )
    \over 1920
  } ,
  \cr
  M^\oneloop(1^-,2^+,3^+,4^+) &= N_s  \Bigl({s t \over
    u} \Bigr)^2\Bigl({ \spb2.4^2 \over \spb1.2 \spa2.3 \spa3.4\spb4.1}
  \Bigr)^2 {( s^2 +st +t^2 ) \over 5760 },
\cr}
\equn\label{eqFourPoints}
$$
where $s=(k_1+k_2)^2$, $t=(k_1+k_4)^2$, $u=(k_1+k_3)^2$ and $N_s =
N_B-N_F$ is the number of bosonic states in the loop minus the number
of fermionic states. The amplitudes for pure gravity are found by
putting $N_s=2$ in the above expressions since a graviton has two
helicity states. These amplitudes vanish in any supersymmetric
theory.
The $n$-point all-plus and one-minus amplitudes are also particle-type independent up to a prefactor of $N_s$,
as can be seen from  the vanishing of these amplitudes in any supersymmetric theory as a consequence of 
supersymmetric Ward identities~\cite{Grisaru:1976vm,Grisaru:1977px}. It is therefore sufficient, for these configurations, to compute the amplitude with a scalar particle circulating in the loop. 

Beyond four points most of the explicit graviton amplitudes are for
scattering in supersymmetric theories.  For $\NeqEight$ supergravity
the $n$-point MHV is known~\cite{Bern:1998sv}, as are the NMHV
six-~\cite{BjerrumBohr:2005xx,Bern:2005bb} and seven-point
amplitudes~\cite{BjerrumBohr:2006yw}.  In ref.~\cite{Bern:1998sv} a
`dimension shift' relation~\cite{Bern:1996ja} allowed the conjecture
of an ansatz for the all-plus $n$-point amplitudes.
This amplitude is an ingredient in the recursion of the one-minus amplitude.

\section{String-based rules calculation of $M^\oneloop(\mpppp)$}
The string-based rules were introduced in 
refs.~\cite{Bern:1991an, Bern:1991aq,Bern:1992cz} as a method of calculating
(one-loop) gauge theory amplitudes.  Their extension to gravity, in
the form we use here, was given in refs.~\cite{Bern:1993wt,
  Dunbar:1994bn}.  In this appendix we summarise these rules
and then describe how they are applied to compute 
$M^\oneloop(1^-,2^+, 3^+, 4^+, 5^+)$. Our presentation treats the method as
something of a `black box' for obtaining field-theory results and we
refer those interested in the details of its string-theory origins and
derivation to the literature.

\subsection{Summary of the string-based rules for gravity amplitudes}
String-based rules use one-loop $\phi^3$-like graphs
to compute the one-loop
corrections to a field theory amplitude.  The terms produced 
take the form of a rational function of the kinematic variables,
within a Feynman parametrisation of a tensor loop integral.
This approach has the advantage of significant computational savings
over the traditional Feynman graph method: far fewer graphs are
involved and early application of simplifications from the
spinor-helicity formalism reduce the complexity of the associated
expressions.

We begin by drawing all one-loop $\phi^3$ graphs excluding massless
bubbles (which vanish in dimensional regularisation) and tadpoles. We
label the outermost legs of the graphs with the particles' momenta,
$k_1, \dots, k_n$. An internal line bears the same label as the first
line or leg found going anti-clockwise about its outermost
vertex. (For examples of such graphs and labellings, see
\figref{fig:sbr-graphs}.) All independent labellings of external legs contribute
for gravity amplitudes.  The one-loop correction to the amplitude is
then given by
\begin{equation*}
  {M}^{\oneloop}(1,2,\dots,n) = 
  \sum_{\text{graphs $\gamma$}} \mathcal{D}(\gamma),
\end{equation*}
where the contribution from a graph $\gamma$ with an
$n_{\ell}$-propagator loop is
\begin{equation}
  \label{eq:sbr-integ}
  \mathcal{D}(\gamma) = \Gamma(n_{\ell} - 2 + \epsilon)
  \left(\prod_{m=1}^{n_{\ell}-1} \int_0^{x_{i_{m+1}}} dx_{i_m} \right)
  \frac{K^{(\gamma)}(x_{i_1},\dots,x_{i_{n_{\ell}-1}})}{
    \left\{ \sum_{1\le k < l \le n_{\ell}} P_{i_k} \cdot P_{i_l}
      x_{i_ki_l} (1 - x_{i_ki_l}) \right\}^{n_{\ell}-2+\epsilon}
  }.
\end{equation} 
In this formula, ${i_1},\dots,{i_{n_{\ell}}}$ are the labels of the
lines adjoining the loop going clockwise.  $x_{ij} \equiv x_i - x_j$,
with $x_{i_{n_{\ell}}}$ fixed at $1$, and $P_{i_k}$ is the momentum
entering the loop along the line with innermost label $i_k$. The
`reduced kinematic factor' for $\gamma$,
$K^{(\gamma)}(x_{i_1},\dots,x_{i_{n_{\ell}-1}})$, is a polynomial in
the $x_{i_k}$ which for a gravity theory with no supersymmetries is of
order $2n_{\ell}$.

We compute $K^{(\gamma)}(x_{i_1},\dots,x_{i_{n_{\ell}-1}})$ as
follows: the starting point is the overall graviton kinematic factor
\begin{multline}
  \label{eq:sbr-kinematicK}
  \mathcal{K} = \int \prod_{i=1}^n dx_i d\bar x_i \prod_{1\le i<j\le
    n} \exp \left\{k_i\cdot k_j \, G^{ij}_{\text{B}} + (k_i \cdot
    \varepsilon_j - k_j \cdot \varepsilon_i) \dot G^{ij}_{\text{B}} -
    \varepsilon_i \cdot \varepsilon_j \, \ddot
    G^{ij}_{\text{B}}\right. \\ \left.\left.  + (k_i \cdot
      \bar\varepsilon_j - k_j \cdot \bar\varepsilon_i) \dot{\bar
        G}{}^{ij}_{\text{B}} - \bar\varepsilon_i \cdot
      \bar\varepsilon_j \, \ddot{\bar G}{}^{ij}_{\text{B}} -
      (\varepsilon_i \cdot \bar\varepsilon_j - \varepsilon_j \cdot
      \bar\varepsilon_i) H^{ij}_{\text{B}}
    \right\}\right|_{\text{multi-linear}},
\end{multline}
where `multi-linear' indicates that we retain only the coefficient of
$\prod_{i=1}^n \varepsilon_i \bar\varepsilon_i$ in the expansion of
the exponentials. The graviton polarisation tensor are then reconstructed using
\eqref{eq:gravpol}. $\mathcal{K}$ contains much structure from the
string theory perspective: $G_{\text B}^{ij} \equiv G_{\text
  B}(x_{ij})$ is the bosonic Green's function on the string
world-sheet, and the $x_i$($\bar x_i$) are closed-string
left(right)-moving co-ordinates. Other objects present are the
derivatives of $G_{\text B}^{ij}$: $\dot G^{ij}_{\text{B}} = \partial
G_{\text B}^{ij} / \partial x_i$, which is antisymmetric in $i,j$, and
$\ddot{ G}^{ij}_{\text{B}} = \partial^2 G_{\text B}^{ij} / \partial
x_i^2$ (with similar expressions for the right-moving $\dot{\bar
  G}^{ij}_{\text{B}}$ and $\ddot{\bar G}^{ij}_{\text{B}}$); and
$H^{ij}_{\text{B}} = \partial^2 G_{\text{B}}^{ij} / \partial
x_i \partial \bar x_i$. However, for our purposes we will simply treat
\eqref{eq:sbr-kinematicK} as an object for obtaining reduced kinematic
factors by the application of some substitution rules that implement
the field theory limit.

For the helicity configuration under consideration and a judicious
choice of reference momenta for the polarisation vectors, we shall see
that the coefficients of the second-order derivatives of the Green's
functions vanish. Nevertheless, in general this is not so and we
should eliminate the $\ddot G_{\text B}$ and $\ddot{\bar G}_{\text B}$
from $\mathcal{K}$ using integration by parts, which may lead to
additional factors of the $H_{\text B}$ appearing. Each $H_{\text B}$
factor should then be eliminated by replacing it with the Feynman
denominator relevant to the diagram under consideration (\ie\ the
expression found within the curly braces in \eqref{eq:sbr-integ}). At
this point we simply drop the integration over the world-sheet
co-ordinates and the leading factors of $\exp(k_i\cdot k_j\,
G^{ij}_{\text{B}})$ from $\mathcal{K}$ (their contributions are built
into the rules).

Now consider a consecutive pair of lines joining on to the loop in a
$\phi^3$ graph, labelled $(i,j)$ going clockwise. We can `pinch off'
this pair of lines by attaching them instead to a new vertex and then
drawing a new line carrying the label $j$ from this vertex back to the
loop. Any of the graphs drawn for string-based rules may be obtained
this way (for example, the graph of \figref{fig:sbr-graphs-lobster} is
obtained from \figref{fig:sbr-graphs-pent} first by pinching off
$(2,3)$, and then $(3,4)$).  For each such pinch used to reach a graph
we: (1) discard all terms in the expression obtained above
\emph{except} those containing \emph{exactly} one power of $\dot
G^{ij}_{\text{B}} \dot{\bar G}{}^{ij}_{\text{B}}$; (2) replace $i$
with $j$ in all remaining $\dot G_{\text B}$, $\dot{\bar G}_{\text
  B}$; and (3) multiply by $-1/k_{ij}^2$, where $k_{ij}$ is the
momentum carried by the new line formed by the pinch.

Next we apply the substitution rules. These act on the derivatives of
the Green's functions in a left/right-independent manner, replacing
them with polynomials in the $x_{ij}$ in a way that depends on the
particle content of the loop. 
In particular, the rule for a single scalar degree of freedom
running around the loop is the simple substitution
\begin{equation}
  \label{eq:sbr-SubGrnS}
  \dot G^{ij}_{\text{B}}, \dot{\bar  G}{}^{ij}_{\text{B}} \rightarrow
  x_{ij} - \frac12 \operatorname{sign} x_{ij}.
\end{equation}
There are other rules for particles of higher spin in the loop
(including the graviton), but
by the discussion in \appref{sec:gravscat},
\eqref{eq:sbr-SubGrnS} is all we need for a
one-minus amplitude.
Therefore we compute the amplitude by
applying \eqref{eq:sbr-SubGrnS} to the reduced kinematic factors and
multiplying by $N_s=2$.
Finally we make
change of the integration variables in \eqref{eq:sbr-integ} to $a_k$
using $x_{i_k} = \sum_{l=1}^k a_l$, which yields an integral in the usual
Feynman parametrisation.

\subsection{Application to $M^\oneloop(1^-,2^+,3^+,4^+,5^+)$}
\begin{figure}[h!]
  \centering\subfloat[]{
    \begin{picture}(120,100)
      \SetOffset(60,50)
      \CArc(0,0)(20,0,360)
      \Line(20,0)(40,0)
      \Line(6.1803,19.0211)(12.3607,38.0423)
      \Line(-16.1803,11.7557)(-32.3607,23.5114)
      \Line(-16.1803,-11.7557)(-32.3607,-23.5114)
      \Line(6.1803,-19.0211)(12.3607,-38.0423)
      \Vertex(20,0){1.5}
      \Vertex(6.1803,19.0211){1.5}
      \Vertex(-16.1803,11.7557){1.5}
      \Vertex(-16.1803,-11.7557){1.5}
      \Vertex(6.1803,-19.0211){1.5}
      \Text(-33,25)[br]{$1$}
      \Text(12.3,40)[cb]{$2$}
      \Text(42,0)[cl]{$3$}
      \Text(12.3,-40)[vt]{$4$}
      \Text(-33,-25)[tr]{$5$}
    \end{picture}
    \label{fig:sbr-graphs-pent}
  }\quad\subfloat[]{
    \begin{picture}(120,100)
      \SetOffset(57.5,50)
      \CArc(0,0)(20,0,360)
      \Line(-20,0)(-40,0)
      \Line(20,0)(35,0)
      \Line(0,20)(0,40)
      \Line(0,-20)(0,-40)
      \Line(35,0)(45,17.3205)
      \Line(35,0)(45,-17.3205)
      \Vertex(20,0){1.5}
      \Vertex(35,0){1.5}
      \Vertex(-20,0){1.5}
      \Vertex(0,20){1.5}
      \Vertex(0,-20){1.5}
      \Text(0,-43)[tc]{$3$}
      \Text(-43,0)[cr]{$4$}
      \Text(0,43)[bc]{$5$}
      \Text(46,20)[bl]{$1$}
      \Text(46,-19)[tl]{$2$}
      \Text(27.5,-2)[tc]{\tiny{$2$}}
    \end{picture}
    \label{fig:sbr-graphs-box}
  }\\ \subfloat[]{
    \begin{picture}(120,100)
      \SetOffset(60,50)
      \CArc(0,0)(20,0,360)
      \Line(0,-20)(0,-40)
      \Line(17.3205,10)(30.3109,17.5)
      \Line(30.3109,17.5)(47.6314,7.5)
      \Line(30.3109,17.5)(30.3109,37.5)
      \Line(-17.3205,10)(-30.3109,17.5)
      \Line(-30.3109,17.5)(-47.6314,7.5)
      \Line(-30.3109,17.5)(-30.3109,37.5)
      \Vertex(0,-20){1.5}
      \Vertex(17.3205,10){1.5}
      \Vertex(-17.3205,10){1.5}
      \Vertex(30.3109,17.5){1.5}
      \Vertex(-30.3109,17.5){1.5}
      \Text(-30.3109,41)[bc]{$2$}
      \Text(30.3109,41)[bc]{$3$}
      \Text(50,9)[tl]{$4$}
      \Text(-50,9)[tr]{$1$}
      \Text(0,-43)[tc]{$5$}
      \Text(24.,13)[tl]{\tiny{$4$}}
      \Text(-24.,15.5)[bl]{\tiny{$2$}}
    \end{picture}
    \label{fig:sbr-graphs-pigtail}
  }\quad\subfloat[]{
    \begin{picture}(120,100)
      \SetOffset(42.5,50)
      \CArc(0,0)(20,0,360)
      \Line(20,0)(50,0)
      \Line(-10,17.3205)(-20,34.6410)
      \Line(-10,-17.3205)(-20,-34.6410)
      \Line(50,0)(60,17.3205)
      \Line(50,0)(60,-17.3205)
      \Line(35,0)(35,-20)
      \Vertex(-10,17.3205){1.5}
      \Vertex(-10,-17.3205){1.5}
      \Vertex(20,0){1.5}
      \Vertex(35,0){1.5}
      \Vertex(50,0){1.5}
      \Text(-22,36.6410)[br]{$5$}
      \Text(35,-23)[tc]{$3$}
      \Text(61,20)[bl]{$1$}
      \Text(61,-19)[tl]{$2$}
      \Text(-22,-36.6410)[tr]{$4$}
      \Text(27.5,2)[bc]{\tiny{$3$}}
      \Text(42.5,2)[bc]{\tiny{$2$}}
    \end{picture}
    \label{fig:sbr-graphs-lobster}
  }
  \caption{%
    Topologies for $\phi^3$-like Feynman diagrams that have a
    non-vanishing contribution to the string-based rules calculation
    of $M^\oneloop (1^-,2^+,3^+,4^+,5^+)$. (The labellings shown are
    non-vanishing examples; other orderings also contribute.)
  }
  \label{fig:sbr-graphs}
\end{figure}
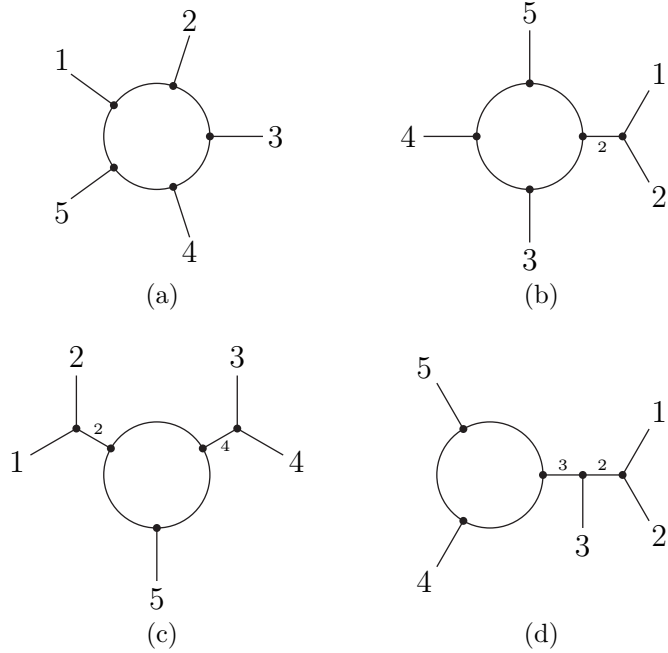

The (topologies of the) graphs that have a non-vanishing contribution
to $M^\oneloop(1^-,\allowbreak2^+,\allowbreak3^+,4^+,5^+)$ are shown
in \figref{fig:sbr-graphs}. There are $117$ such labelled graphs in
total: $12$ massless pentagons (\figref{fig:sbr-graphs-pent}), $30$ one-mass
boxes (\figref{fig:sbr-graphs-box}), $15$ two-mass triangles
(\figref{fig:sbr-graphs-pigtail}) and $30$ one-mass triangles
(\figref{fig:sbr-graphs-lobster}). There are also $30$ massive
bubbles, but these vanish by the pinching process when using the spinor helicity choice~(\ref{eq:poln-spinors}) below. 

In order to define the $\varepsilon_i$ and $\bar\varepsilon_i$ (which
are set to the same values after multi-linearisation in
\eqref{eq:sbr-kinematicK}), we choose $k_5$ as the reference momentum
for the first graviton and $k_1$ for the rest. In the spinor-helicity
formalism the polarisation vectors are
\begin{equation}
  \label{eq:poln-spinors}
  \varepsilon_1^\mu =  \frac{\spba 5.{\gamma^\mu}.1}{\sqrt 2 \spb 1.5}
  \qquad\text{and}\qquad
  \varepsilon_i^\mu =  \frac{\spba i.{\gamma^\mu}.1}
  {\sqrt 2 \spa 1.i}
  \quad\text{for $i \neq 1$}.
\end{equation}
We have the standard spinor-helicity results that $k_i \cdot
\varepsilon_i = k_5 \cdot \varepsilon_1 = k_1 \cdot \varepsilon_i = 0$
for all $i$, and furthermore for this choice $\varepsilon_i \cdot
\varepsilon_j$ vanishes for all $i,j$, so there are no second
derivatives of Green's functions to handle. After 
dropping the $\exp(k_i\cdot k_j G_{\text{B}}^{ij})$,
\eqref{eq:sbr-kinematicK} becomes,

\newcommand{\Gr}[2]{k_{#2}\cdot\varepsilon_{#1}\,\dot
  G_{\text{B}}^{#1#2}}
\begin{multline}
  \label{eq:sbr-amp-K}
  (\Gr12 + \Gr13 + \Gr14)(\Gr23 + \Gr24 + \Gr25) \\
  \times (\Gr32 + \Gr34 + \Gr35)(\Gr42 + \Gr43 + \Gr45) \\
  \times (\Gr52 + \Gr53 + \Gr54) \times (\text{l}\rightarrow\text{r}).
\end{multline}
Here, `$(\text{l}\rightarrow\text{r})$' denotes taking the expression
to the left and replacing all $\dot G_{\text{B}}^{ij}$ with
$\dot{\bar G}_{\text{B}}^{ij}$. We can now see why there are no bubble
graphs in the problem. They come from three pinches that form two
independent trees, but any such sequence of pinches will either pull
out a factor of the form $(k_1+k_i)\cdot\varepsilon_i$, which vanishes
by conservation of momentum and the remarks below
\eqref{eq:poln-spinors}, or simply run out of pinchable $G_{\text{B}}$s.

Since the same substitution rule \eqref{eq:sbr-SubGrnS} is
applied independently to both
the left- and right-moving sectors, and at each step in the
pinching we pull out terms containing exactly one power of both $\dot
G_{\text{B}}^{ij}$ and $\dot{\bar G}_{\text{B}}^{ij}$, we can in fact
proceed in a rather more straightforward manner by applying the
pinching and substituting for just the left-moving factors of
\eqref{eq:sbr-amp-K}, then taking the square of the result as
$K^{(\gamma)}(x_{i_1},\dots,x_{i_{n_{\ell}-1}})$, taking care
\emph{not} to square the kinematic factors that arise from the trees
during pinching.

We do this for all 117 graphs and substitute back into
\eqref{eq:sbr-integ}, changing the variables to the usual Feynman
parameters. Each graph thereby yields an expression of the form
\begin{equation}
  \label{eq:integ-schematic}
  \mathcal{D}(\gamma) = 
  \sum_{\{p_i\}} X^{(\gamma)}(r_1,\dots, r_{n_\ell})
  I^{(\gamma)}_{n_\ell}[a_1^{r_1} \cdots a_{n_\ell}^{r_{n_\ell}}],
\end{equation}
where $X^{(\gamma)}(r_1,\dots, r_{n_\ell})$ is a rational coefficient.
The $n_\ell$-gonal tensor Feynman integral with momentum configuration
relevant to the graph $\gamma$ is defined as
\begin{equation}
  \label{eq:tensorintegral}
  I^{(\gamma)}_{n_\ell}[a_1^{r_1} \cdots
  a_{n_\ell}^{r_{n_\ell}}] = \Gamma(n_\ell - 2 + \epsilon) \int_0^1
  d^{n_\ell}a \frac {a_1^{r_1} \cdots
    a_{n_\ell}^{r_{n_\ell}} \delta(1-\sum_i a_i)}{
    \left\{ \sum_{k,l=1}^{n_\ell} S^{(\gamma)}_{kl}a_k a_l - 
      i \varepsilon \right\}^{n_\ell-2+\epsilon}
  },
\end{equation}
with the array $S^{(\gamma)}_{kl}$ given in terms of the momenta
$P_{i_k}$ entering $\gamma$'s loop by
\[
S^{(\gamma)}_{kl} = \begin{cases}
  0 & \text{for $k=l$},\\
  -\tfrac12 (P_{i_k} + \cdots + P_{i_{l-1}})^2 & \text{otherwise.}
\end{cases}
\]

These integrals may be evaluated by the recursive approach detailed in
refs.~\cite{Bern:1993mq, Bern:1993kr}. In this way we have
constructed an integral database using computer algebra, indexed by
the tuples $\{(r_1,\dots, r_{n_\ell}) | \sum_i r_i \le 2n_\ell\}$ for
integrals with $3 \le n_\ell \le 5$ and up to $5 - n_\ell$ massive
legs. The table contains both the rational pieces of the integrals and
the rational coefficients of their (di)logarithms.  Since
$M^\oneloop(1^-,2^+,3^+,4^+,5^+)$ is entirely rational, we can use the
vanishing of the logarithmic terms as a consistency check.
Computationally the most complicated integrals are the pentagons where we must evaluate integrals with Feynman parameter polynomials of order ten.

The
coefficients produced in this manner are far too large and cumbersome
to present here (or even form a compact analytic expression for the
amplitude at present); nevertheless, they are amenable to exact
numeric arithmetic at a kinematic
point.  The results have been compared  with, and agree with, 
the recursion-derived expression \eqref{eq:recursion-amp}.

\end{document}